\documentclass[aps,prb,twocolumn,superscriptaddress,nofootinbib,floatfix]{revtex4-2}
\usepackage[utf8]{inputenc}
\usepackage[english]{babel}
\usepackage[expansion,protrusion]{microtype}
\usepackage{amsmath}
\usepackage{amssymb}
\usepackage{xfrac}
\usepackage[normalem]{ulem}
\usepackage{array}
\usepackage{siunitx}
\usepackage{graphicx}
\usepackage[dvipsnames]{xcolor}
\usepackage[colorlinks=true,citecolor=blue,linkcolor=blue,urlcolor=blue]{hyperref}
\usepackage{booktabs}
\usepackage{multirow}
\usepackage{orcidlink}
\newcolumntype{C}[1]{>{\centering\arraybackslash}m{#1}}

\useshorthands*{"}
\defineshorthand{"=}{\babelhyphen{hard}}
\DeclareMathOperator\real{Re}
\DeclareMathOperator\imag{Im}
\DeclareSIUnit\Ry{Ry}

\begin{document}

\title{Self-doped correlated adatom lattices on cubic boron nitride}

\author{Niklas~Enderlein\,\orcidlink{0000-0001-9832-1109}}
\email[Contact author: ]{niklas.h.enderlein@fau.de}
\affiliation{Department of Physics, Friedrich-Alexander-Universit\"at Erlangen-N\"urnberg, 91058 Erlangen, Germany}

\author{Tim~Kullick\,\orcidlink{0009-0002-9956-4692}}
\affiliation{Department of Physics, Friedrich-Alexander-Universit\"at Erlangen-N\"urnberg, 91058 Erlangen, Germany}

\author{Henri~Menke\,\orcidlink{0000-0001-5245-5400}}
\affiliation{Department of Physics, Friedrich-Alexander-Universit\"at Erlangen-N\"urnberg, 91058 Erlangen, Germany}
\affiliation{Max Planck Computing and Data Facility, 85748 Garching b.\ M\"unchen, Germany}

\author{Roland~Gillen\,\orcidlink{0000-0002-7913-0953}}
\affiliation{Department of Physics, Friedrich-Alexander-Universit\"at Erlangen-N\"urnberg, 91058 Erlangen, Germany}
\affiliation{College of Engineering, Swansea University, Swansea SA1 8EN, United Kingdom}

\author{Janina~Maultzsch\,\orcidlink{0000-0002-6088-2442}}
\affiliation{Department of Physics, Friedrich-Alexander-Universit\"at Erlangen-N\"urnberg, 91058 Erlangen, Germany}

\author{Philipp~Hansmann\,\orcidlink{0000-0002-0330-7927}}
\email[Contact author: ]{philipp.hansmann@fau.de}
\affiliation{Department of Physics, Friedrich-Alexander-Universit\"at Erlangen-N\"urnberg, 91058 Erlangen, Germany}
\affiliation{Science Institute and Faculty of Physical Sciences, University of Iceland, 107 Reykjav\'ik, Iceland}

\date{\today}

\begin{abstract}
We survey triangular adatom lattices on the (111) surfaces of cubic boron nitride (cBN) for strongly correlated two-dimensional electron systems.
Using density functional theory combined with the constrained random phase approximation and charge self-consistent dynamical mean-field theory, we identify twelve promising candidates.
Group-III (group-V) adatoms on the N- (B-)termination realize clean one-band Hubbard systems, while group-III and transition-metal (V, Cr, Mn) adatoms on the B-termination realize two-band Hubbard systems.
Importantly, the polar III–V character of cBN induces self-doping by $\pm\sfrac{1}{4}$ electron away from integer filling.
With strong Hubbard (and Hund's) interactions and small hoppings, most systems are doped Mott insulators.
Heavy quasielectrons with mass renormalizations up to $4$ and strong local moments up to $2.5\,\mu_B$ put our systems into a regime where unconventional superconductivity on the triangular lattice has been predicted.

\end{abstract}

\maketitle

\section{Introduction}
\label{section:intro}
Adatom lattices on semiconductor surfaces are a versatile platform for strongly correlated electron physics in two dimensions.
Group-IV adatoms (Sn, Pb, C) on Si(111) and Ge(111) at $\sfrac{1}{3}$ monolayer coverage have been shown to realize (extended) one-band Hubbard models on triangular lattices~\cite{carpinelli1996direct,carpinelli1997surface,profeta2005novel,profeta2007triangular,schuwalow2010realistic,li2011geometrical,hansmann2013jpcm,hansmann2013prl,gangli2013, Vandelli2024, Bunney2025, Torkzadeh2026}.
Their surface bands are narrow, so that the effective Coulomb interaction is sizable compared to the kinetic energy scale, which leaves them in a correlated regime that is far from either perturbative limit.
In this regime, strongly renormalized quasielectrons, Mott insulating behavior, and frustrated magnetism of localized moments have been studied.

Beyond these, chiral $d$-wave superconductivity has been predicted for hole-doped Sn/Si(111)~\cite{cao2018} and was subsequently observed~\cite{weitering2020}, with experimental evidence for chiral pairing~\cite{weitering2023}. Further studies of the doped triangular-lattice $t$-$J$ model support a chiral order parameter~\cite{huang2022topological,huang2023chiral}, although its specific symmetry is still under debate~\cite{wolf2022triplet}. The pairing mechanism is likely electronic, as electron-phonon coupling appears too weak to account for the observed critical temperature~\cite{marchetti2025phonons}.
To reach the superconducting state, however, a relatively high amount of hole doping was key~\cite{cao2018,weitering2020}. Realizing such high doping is hard without introducing simultaneously detrimental effects such as disorder and surface degradation~\cite{weitering2020} and platforms which allow for a ``cleaner'' inception of the doped charge are highly desirable.

The substrate is an obvious lever to tailor both the bandwidths and the fillings of adatom lattices.
For the same adatom species, replacing Si with SiC yields narrower bands and a larger effective Coulomb interaction: the short, strong covalent Si-C bonds open a wide band gap, which places the adatom state far from the substrate bands.
The surface orbital therefore leaks less into the substrate, which reduces the effective hopping and increases the bare interaction, while the low polarizability of the wide-gap substrate weakens the screening~\cite{hansmann2013jpcm,glass2015triangular}.
Following this route, we found in a recent study~\cite{menke2025} that transition-metal (TM) adatoms (Ti, V, Cr) on 3C-SiC(111) realize correlated two-band Hubbard models with Dirac cones and flat bands, with Mott insulating and heavy Fermi-liquid ground states. The fillings of these lattices, however, stay at the integer values set by the group-IV substrate.

In this work, we extend our materials exploration to cubic boron nitride (cBN), a III-V compound semiconductor in the zincblende structure, which offers two advantages over SiC. First, its wider band gap (${\sim\SI{6.4}{\eV}}$ vs.~${\sim\SI{2.4}{\eV}}$ for 3C-SiC) and smaller dielectric constant further enhance the Coulomb interaction and suppress the effective hopping.
Second, and crucially, cBN(111) is a polar surface~\cite{tasker1979stability} in which, unlike the isovalent Si and C in SiC, B and N contribute unequal numbers of valence electrons per bond ($\sfrac{3}{4}$ for B vs.\ $\sfrac{5}{4}$ for N), so an adatom must supply a different number of electrons to saturate the dangling covalent bonds than on a group-IV substrate~\cite{pashley1989electron}.
The adatom bands are therefore self-doped by ${\pm\sfrac{1}{4}}$ electron per site away from integer filling on the B-~(N-)termination.
In the spirit of modulation doping in semiconductor heterostructures, the doping charge is thus supplied by the substrate and not by dopants in the adatom layer.
Given the strong interaction, such self-doped lattices are natural candidates for doped Mott insulators, the regime in which unconventional superconductivity on the triangular lattice has been predicted and found~\cite{cao2018,weitering2020,weitering2023}.

So far, III-V semiconductor (111) surfaces have received little attention as substrates for correlated adatom lattices.
While GaAs(111) has been used to engineer adatom-induced reconstructions, e.g.\ bismuth on GaAs(111)~\cite{gaas2023bismuth}, the resulting honeycomb structures fall outside the correlated electron regime.
For cBN(111) even less preliminary work exists and the surface -- even without adatoms -- is largely unexplored. Studies on cBN(111) are almost exclusively theoretical~\cite{kadas1998ab,yang2005ab,sun2020structural}, and the experimental characterization is limited to early electron-diffraction, spectroscopy, and atomic-force-microscopy work confirming an unreconstructed $1\times1$ periodicity~\cite{loh1997surface,takami1998domains}.
At the same time, cBN is similar to SiC, for which a growing family of SiC-supported monolayers, from the correlated Sn/SiC(0001) lattice~\cite{glass2015triangular} to the topological indenene~\cite{bauernfeind2021design} and bismuthene~\cite{reis2017bismuthene,gehrig2025bismuthene} systems, has been realized experimentally.

We surveyed adatom species up to the fourth period for stable lattices at $\sfrac{1}{3}$~ML coverage in the established $(\sqrt{3}\times\sqrt{3})R30^\circ$ surface cell on both the B- and the N-termination of cBN(111).
This resulted in twelve candidate adatom lattices, for which we combine density functional theory (DFT) with the constrained random phase approximation and fully charge self-consistent dynamical mean-field theory (DMFT).
Six systems realize one-band Hubbard models: group-III adatoms (B, Al, Ga) on the N-termination, and group-V adatoms (N, P, As) on the B-termination. The remaining six realize two-band Hubbard models, formed by group-III (B, Al, Ga) and $3d$ TM (V, Cr, Mn) adatoms, all on the B-termination.
In all systems except Al and Ga on the B-termination, which form correlated Fermi liquids, the interaction exceeds the bandwidth by more than a factor of four ($U/W>4$), so that the self-doped adatom lattices realize doped Mott insulators above the Fermi liquid coherence temperature.
For the TM adatoms, we find Dirac cones and flat bands close to the Fermi level, similar to our findings for 3C-SiC(111)~\cite{menke2025}.

The paper is organized as follows.
Section~\ref{section:method} describes the slab models and computational methods.
In Sec.~\ref{section:dft}, we discuss the adsorption sites, band structures, fillings and interaction parameters.
Section~\ref{section:dmft} presents the DMFT results for the one- and two-band systems
and, for Cr on the B-termination, follows the filling from the Mott insulator at integer filling to the self-doped metallic state.
Section~\ref{section:conclusions} summarizes our results and gives an outlook.
\section{Methods and Models}
\label{section:method}
\paragraph{Construction of the slab geometries.}
We constructed slab geometries based on DFT-optimized bulk cubic boron nitride (cBN) in the zincblende structure, exposing the (111) surface with both B- and N-terminations, using a slab thickness of eight BN bilayers. Adatoms at $\sfrac{1}{3}$~ML coverage form a $\sqrt{3} \times \sqrt{3}$ lattice on the respective surface termination. Both adsorption geometries, $\mathrm{T}_4$ and $\mathrm{H}_3$, were relaxed for each system. The choice of site is discussed in Sec.~\ref{subsection:slab}. Due to the III--V character of cBN, the opposite slab surface cannot be fully passivated with standard hydrogen atoms and instead requires one pseudo-hydrogen per terminating N or B atom, carrying fractional charges of 0.75 or 1.25 electrons, respectively, as employed previously~\cite{yang2005ab,sun2020structural}. A vacuum spacing of \SI{22}{\angstrom} minimized inter-slab interactions. In-plane lattice constants were fixed to the DFT-optimized bulk values. Atomic positions were optimized for the adatoms and the top three substrate layers, while deeper layers were fixed to simulate the semi-infinite bulk. Structural optimizations were performed with VASP~\cite{kresse1993ab,kresse1996efficiency,kresse1996efficient}, using the PBE exchange-correlation functional, an \SI{800}{\eV} energy cutoff and $12 \times 12 \times 1$ $\mathbf{k}$-point meshes, until residual forces fell below $\SI[per-mode=symbol]{e-3}{\eV\per\angstrom}$.

\paragraph{cRPA}
We define the correlated subspace in a Wannier basis constructed with Wannier90~\cite{pizzi2020wannier90}, obtained by downfolding the adatom-associated bands at the Fermi energy $E_\mathrm{F}$, i.e., one apical $p_z$-like orbital for the one-band systems and an $e$ doublet of planar-$p$ or $3d$ character for the two-band systems (Sec.~\ref{section:dft}). This basis was used both for the computation of the static screened Coulomb interaction within the constrained random phase approximation (cRPA)~\cite{aryasetiawan2004frequency,vaugier2012hubbard} and as the projector basis for the DFT+DMFT up-/downfolding. For the cRPA calculations we used the VASP~\cite{kresse1993ab,kresse1996efficiency,kresse1996efficient} implementation, employing $GW$-type pseudopotentials, a \SI{600}{\eV} energy cutoff, $12 \times 12 \times 1$ $\mathbf{k}$-point meshes, and 610 empty bands (corresponding to states up to $\approx$\SI{50}{\eV} above $E_\text{F}$). The resulting interaction parameters were used to construct a Hubbard-Kanamori Hamiltonian for the DMFT calculations,
\begin{equation}
\begin{split}
\hat{H}_{\mathrm{int}} = &\ U \sum_m \hat{n}_{m\uparrow} \hat{n}_{m\downarrow}
+ U' \sum_{m\neq m'} \hat{n}_{m\uparrow} \hat{n}_{m'\downarrow} \\
&+ (U' - J_{\mathrm{H}}) \sum_{m < m', \sigma} \hat{n}_{m\sigma} \hat{n}_{m'\sigma} \\
&- J_{\mathrm{H}} \sum_{m \neq m'} \hat{c}^\dagger_{m\uparrow} \hat{c}^{\phantom{\dagger}}_{m\downarrow}
\hat{c}^\dagger_{m'\downarrow} \hat{c}^{\phantom{\dagger}}_{m'\uparrow} \\
&+ J_{\mathrm{H}} \sum_{m \neq m'} \hat{c}^\dagger_{m\uparrow} \hat{c}^\dagger_{m\downarrow}
\hat{c}^{\phantom{\dagger}}_{m'\downarrow} \hat{c}^{\phantom{\dagger}}_{m'\uparrow}
\end{split}
\end{equation}
with intra-/inter-orbital interaction $U/U'$ and Hund's coupling $J_{\mathrm{H}}$. For the one-band systems, $\hat{H}_{\mathrm{int}}$ reduces to the single-orbital Hubbard term $U \hat{n}_{\uparrow}\hat{n}_{\downarrow}$.

\paragraph{Charge self-consistent DFT+DMFT}
We performed fully charge self-consistent DFT+DMFT~\cite{Anisimov1997, Lichtenstein1998, dmft3} using the \textsc{solid\_dmft} package~\cite{Merkel2022, Beck2022}, interfacing Quantum ESPRESSO~\cite{giannozzi2009quantum,giannozzi2017advanced} (DFT) and the TRIQS library~\cite{Parcollet2015} (DMFT). The DFT part of the self-consistency loop used norm-conserving pseudopotentials~\cite{hamann2013optimized,van2018pseudodojo} with an energy cutoff of \SI{120}{\Ry} and a $12 \times 12 \times 1$ $\mathbf{k}$-point mesh.
Starting from a converged DFT ground state, DFT and DMFT calculations were iterated until convergence of the DMFT self-energy, using TRIQS/\textsc{cthyb}~\cite{Werner2006, Gull2011, Seth2016} at an inverse temperature $\beta = \SI{100}{\per\eV}$, with $10^6$ warm-up and $10^7$ Monte Carlo cycles with a cycle length of 200 steps each.
Green's functions were represented on 1000 Matsubara frequencies and 6001 imaginary-time points, and 30 Legendre coefficients were used to improve convergence. Real-frequency spectral functions were obtained from the Matsubara-axis QMC data by analytic continuation using the Maximum Entropy method~\cite{bergeron2016algorithms}. Error bars on the resulting self-energies and spectral functions, shown as shaded areas in the corresponding plots, were estimated by exploiting the residual dependence of the analytically continued self-energy on the auxiliary parameter $\mu_{\mathrm{aux}}$ (see Supplemental Material~\cite{Supplemental}).

Spin-orbit coupling, which would in principle gap the Dirac crossings of the two-band systems (Sec.~\ref{section:dft}), is negligible for the light adatoms and energy scales considered here and was therefore omitted. All DFT calculations were spin-unpolarized.

\paragraph{Phonons}
As we propose these adatom lattices as candidate platforms for future experimental realization, we assessed their dynamical stability within density functional perturbation theory~\cite{baroni2001phonons} as implemented in Quantum ESPRESSO~\cite{giannozzi2009quantum,giannozzi2017advanced}. Phonon dispersions and further details are given in the Supplemental Material~\cite{Supplemental}.

\paragraph{AI tools}
The AI coding assistant Claude Code (Anthropic, model Claude Opus 5.5) was used to optimize the analytic-continuation workflow, i.e., the choice of $\mu_{\mathrm{aux}}$ and of the regularization parameter, to extract the DMFT quantities in Table~\ref{tab:overview} and the tight-binding parameters of the Wannier Hamiltonians (Supplemental Material~\cite{Supplemental}), and to assist with the plotting code for all figures of this article and of the Supplemental Material.

\section{Results: DFT and cRPA}
\label{section:dft}
Table~\ref{tab:overview} summarizes the central results of all twelve adatom systems considered in this work and serves as an overview reference.
In the following, we denote an adatom X on the B- (N-) termination by X@B (X@N).

\subsection{Adsorption sites and stability}
\label{subsection:slab}
\begin{figure*}[p]
\includegraphics[width=\textwidth]{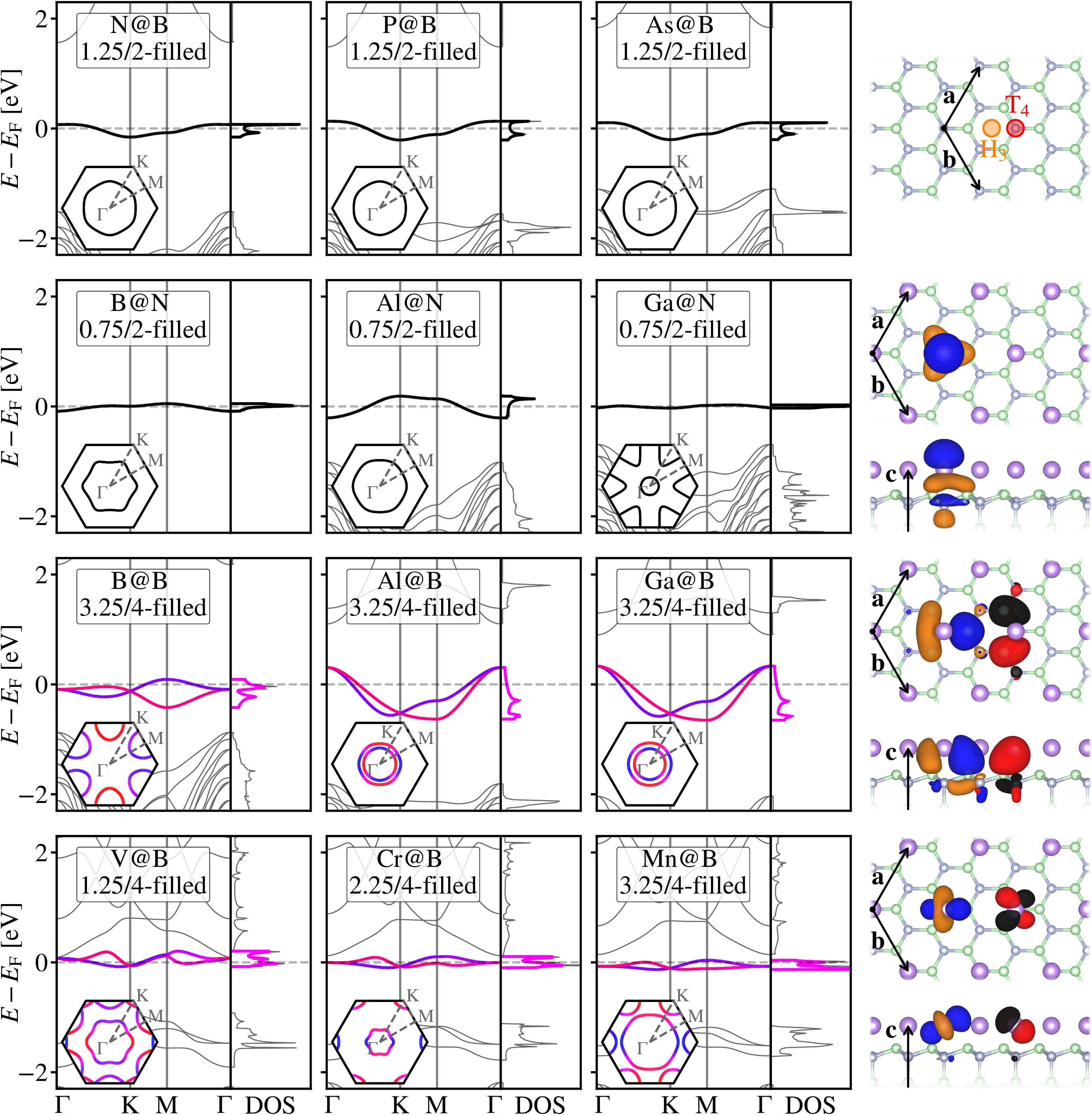}
\caption{DFT band structures of the twelve candidate adatom lattices on cBN(111) along $\Gamma$-$K$-$M$-$\Gamma$ in the hexagonal surface Brillouin zone, with the density of states in the adjacent panel and the Fermi surface as an inset. Thick lines mark the adatom bands of the correlated subspace, thin gray lines all other bands of the slab. Rows from top to bottom: group-V adatoms (N, P, As) on the boron surface, group-III adatoms (B, Al, Ga) on the nitrogen surface, group-III adatoms (B, Al, Ga) on the boron surface, and $3d$ TM adatoms (V, Cr, Mn) on the boron surface. The first two rows are the one-band systems, the last two the two-band systems, for which the color scale from red to blue gives the weight of the two Wannier orbitals shown on the right in the same colors. The label in each panel gives the filling of the adatom manifold. Top right: top view of the cBN(111) surface with the lattice vectors $\mathbf{a}$ and $\mathbf{b}$ of the $(\sqrt{3}\times\sqrt{3})R30^\circ$
cell and the two adsorption sites T$_4$ (red) and H$_3$ (orange). Below, the right column shows the Wannier functions of one representative system per class
(isosurfaces in top and side view, with the two colors giving the sign and large violet spheres marking the adatoms), a single apical $p_z$-like orbital for the one-band systems and two orbitals for the two-band systems.
}
\label{fig:dft_all}
\end{figure*}

At a coverage of $\sfrac{1}{3}$~ML, i.e., one adatom per three top-layer atoms, the zincblende (111) surface offers two high-symmetry adsorption sites at which the adatom forms three equivalent covalent bonds and thereby saturates the dangling bonds of all three top-layer atoms in the $(\sqrt{3}\times\sqrt{3})R30^\circ$ cell: the T$_4$ site directly above a second-layer atom and the H$_3$ hollow site above a fourth-layer atom (Fig.~\ref{fig:dft_all}, top right).
To determine the preferred site, we performed structural relaxations for both adsorption geometries of each system, as described in Sec.~\ref{section:method}, and compared the total energies (Supplemental Material~\cite{Supplemental}) of the resulting optimized structures.
For almost all systems, T$_4$ is favored by at least 0.3~eV per $(\sqrt{3}\times\sqrt{3})R30^\circ$ cell (i.e., per adatom). The only exceptions occur on the N-termination:
Al@N prefers H$_3$ by 0.1~eV, whereas for Ga@N both sites are practically degenerate and yield essentially the same electronic structure. Since the choice of site is irrelevant for Ga@N, we employ for both Al@N and Ga@N the H$_3$ site and use the T$_4$ site in all other cases.
Only this one respective adsorption geometry is considered for each system in the remainder of this work.
The absence of imaginary phonon modes indicates that all adatom lattices are dynamically stable, with the sole exception of B@B, where we find a doubly degenerate unstable mode at the $K$ point, at \SI{-170}{\per\centi\meter}.
Displacing the adatoms along this mode leads to an energetically favored surface reconstruction (see Supplemental Material~\cite{Supplemental}).

\subsection{DFT band structure and Wannier projection}
In Fig.~\ref{fig:dft_all} we plot the DFT bands along the high-symmetry path $\Gamma$-$K$-$M$-$\Gamma$ in the hexagonal Brillouin zone, together with the resulting Fermi surfaces (insets) and density of states (DOS) for all twelve adatom systems.
Since the Wannier orbitals are qualitatively identical within each class of systems, the right column of the figure shows them for only one representative per class.

All twelve systems feature narrow (bandwidth $W\leq\SI{1}{\eV}$) adatom-derived bands around the Fermi level ($E_\mathrm{F}$) that are energetically well isolated from the electronic states of the cBN slab, as the DOS panels show. In the $C_{3v}$ symmetry of the T$_4$ (or H$_3$) adsorption site, the adatom $p$ shell splits into a non-degenerate $a_1$ state and a doubly degenerate $e$ state.
For the one-band systems, the associated Wannier orbital is predominantly of apical $p_z$-like character (see Fig.~\ref{fig:dft_all} row~2), corresponding to the $a_1$ state, the same orbital character hosted by the experimentally realized group-IV adatoms on group-IV substrates~\cite{carpinelli1996direct,carpinelli1997surface,profeta2005novel,profeta2007triangular,schuwalow2010realistic,li2011geometrical,hansmann2013jpcm,hansmann2013prl,gangli2013, Vandelli2024, Bunney2025, Torkzadeh2026}. For the two-band systems formed by group-III adatoms (B, Al, Ga) on the B-termination, by contrast, the associated Wannier orbitals are of planar-$p$ character (see Fig.~\ref{fig:dft_all} row~3), corresponding to the $e$ doublet.
For the TM systems (V, Cr, Mn), the two associated Wannier orbitals originate mainly from the adatom $3d$ shell and likewise form an $e$ doublet, resembling tilted $3z^2{-}\mathbf{r}^2$ and $x^2{-}y^2$ orbitals (see Fig.~\ref{fig:dft_all} row~4), as found for the analogous TM systems on 3C-SiC(111)~\cite{menke2025}.

While all Wannier orbitals contain contributions from the underlying substrate atoms, they are well localized and show spatial extents (square root of the Wannier spread) ranging from 1.7 to \SI{2.9}{\angstrom}, well below the adatom spacing of ${\sim}\SI{4.4}{\angstrom}$.

As for the TM adatoms on 3C-SiC(111)~\cite{menke2025}, all two-band systems exhibit Dirac cones at the $K$ point. Along the $\Gamma$-$K$ and $K$-$M$ lines, the little group in reciprocal space is $C_s$, and the two adatom bands transform according to the irreducible representations $A'$ and $A''$, of even and odd parity under the vertical mirror plane $\sigma_v$ containing the nearest-neighbor adatom bonds (Supplemental Material~\cite{Supplemental}). This mirror symmetry protects the Dirac points in the absence of spin-orbit coupling~\cite{Eck2022}.
In the real-space Wannier basis, the two adatom bands generally have mixed orbital character (color coding in Fig.~\ref{fig:dft_all}), which varies along the $\mathbf{k}$-path with the orientation of $\mathbf{k}$ relative to the Wannier orbitals. Owing to the twofold degeneracy, this composition is not unique but depends on the choice of local reference frame.
\makeatletter\@booleanfalse\force@deferlist@sw\makeatother
\begin{table*}[t]
\centering
\caption{Overview of all adatom systems on cBN(111): electron filling per number of orbitals, Wannier bandwidth ($W$), extent of the Wannier function ($\sqrt{\text{Wannier spread}}$), cRPA interaction parameters ($U$, $U'$, $J_\mathrm{H}$) and DMFT results at $\beta=\SI{100}{\per\eV}$: the effective mass renormalization $Z^{-1}$ from the slope of $\real\Sigma(\omega)$ at $E_\mathrm{F}$, the position of the satellite maximum (not the mean energy, which is discussed in the Supplemental Material~\cite{Supplemental}) of the local spectral function relative to $U$ (below\,/\,above $E_\mathrm{F}$ for three-peak spectra), the scattering rate at $E_\mathrm{F}$ relative to the bandwidth ($-\imag\Sigma(0)/W$) as a measure of quasiparticle coherence and the local moment $m_\text{loc} \simeq \sqrt{3\,k_\text{B}T\, \chi_\text{loc}(\omega{=}0)}$.
Error bars of $Z^{-1}$ and $-\imag\Sigma(0)/W$ denote the standard deviation over the auxiliary chemical potentials used in the analytic
continuation, the same convention as for the envelopes in Figs.~\ref{fig:dmft_spectral} and \ref{Fig-sigma}.
}

\sisetup{separate-uncertainty=true}
  \begin{tabular*}{\linewidth}{@{\extracolsep{\fill}}l c c c S[table-format=1.3] S[table-format=1.3] S[table-format=1.3] S[table-format=1.1(2)] c
  S[table-format=1.2(2)] S[table-format=1.2] @{}}
  \toprule
  system & filling & $W$ [\si{\eV}] & $\sqrt{\text{sprd}}$ [\si{\angstrom}] & {$U$ [\si{\eV}]} & {$U^{\prime}$ [\si{\eV}]} & {$J_{\mathrm{H}}$ [\si{\eV}]} &
  {$Z^{-1}$} & {$\frac{|\omega_\text{sat}|}{U}$} & {$\frac{-\imag\Sigma(0)}{W}$} & {$m_\text{loc}~[\mu_B]$} \\
  \midrule
  \multicolumn{11}{@{}l@{}}{\textit{Group-V adatoms on boron surface (one $p$-orbital)}} \\
  \addlinespace[\aboverulesep]
  N@B  & \sfrac{1.25}{2} & 0.23 & 1.73 & 4.35 & {---} & {---} & 2.5 \pm 0.5 & 0.97 & 0.11 \pm 0.03 & 1.17 \\
  P@B  & \sfrac{1.25}{2} & 0.34 & 2.15 & 3.13 & {---} & {---} & 3.4 \pm 0.7 & 0.97 & 0.06 \pm 0.02 & 1.10 \\
  As@B & \sfrac{1.25}{2} & 0.32 & 2.25 & 2.83 & {---} & {---} & 3.4 \pm 0.7 & 0.97 & 0.07 \pm 0.03 & 1.14 \\
  \midrule
  \multicolumn{11}{@{}l@{}}{\textit{Group-III adatoms on nitrogen surface (one $p$-orbital)}} \\
  \addlinespace[\aboverulesep]
  B@N  & \sfrac{0.75}{2} & 0.14 & 1.83 & 4.23 & {---} & {---} & 1.6 \pm 0.3 & 0.97 & 0.10 \pm 0.03 & 1.32 \\
  Al@N & \sfrac{0.75}{2} & 0.41 & 2.45 & 2.81 & {---} & {---} & 3.5 \pm 0.7 & 0.97 & 0.04 \pm 0.02 & 1.04 \\
  Ga@N & \sfrac{0.75}{2} & 0.05 & 2.87 & 1.92 & {---} & {---} & 1.4 \pm 0.2 & 0.96 & 0.18 \pm 0.05 & 1.45 \\
  \midrule
  \multicolumn{11}{@{}l@{}}{\textit{Group-III adatoms on boron surface (two $p$-orbitals)}} \\
  \addlinespace[\aboverulesep]
  B@B  & \sfrac{3.25}{4} & 0.51 & 1.98 & 3.20 & 2.65 & 0.27 & 4.1 \pm 0.7 & 0.86 & 0.07 \pm 0.02 & 1.05 \\
  Al@B & \sfrac{3.25}{4} & 0.96 & 2.23 & 2.56 & 2.08 & 0.24 & 2.5 \pm 0.3 & {$0.86\,/\,0.18$} & {$<0.01$} & 0.58 \\
  Ga@B & \sfrac{3.25}{4} & 1.00 & 2.21 & 2.58 & 2.11 & 0.23 & 2.4 \pm 0.3 & {$0.87\,/\,0.19$} & {$<0.01$} & 0.55 \\
  \midrule
  \multicolumn{11}{@{}l@{}}{\textit{Transition-metal adatoms on boron surface (two $d$-orbitals)}} \\
  \addlinespace[\aboverulesep]
  V@B  & \sfrac{1.25}{4} & 0.28 & 2.83 & 1.33 & 0.84 & 0.24 & 2.7 \pm 0.9 & {$0.39\,/\,0.63$} & 0.25 \pm 0.05 & 1.85 \\
  Cr@B & \sfrac{2.25}{4} & 0.20 & 2.12 & 1.55 & 0.91 & 0.32 & 2.3 \pm 0.8 & 0.94 & 0.22 \pm 0.05 & 2.47 \\
  Mn@B & \sfrac{3.25}{4} & 0.16 & 1.76 & 1.79 & 1.01 & 0.39 & 2.9 \pm 1.2 & 0.53 & 0.19 \pm 0.11 & 1.26 \\
  \bottomrule
  \end{tabular*}

\label{tab:overview}
\end{table*}
\subsection{Self-doping by the polar surface}
The fillings listed in Table~\ref{tab:overview} follow from a simple electron count~\cite{pashley1989electron}.
We take the experimentally realized group-IV adatoms on group-IV substrates~\cite{carpinelli1996direct,carpinelli1997surface,profeta2005novel,profeta2007triangular,schuwalow2010realistic,li2011geometrical,hansmann2013jpcm,hansmann2013prl,gangli2013, Vandelli2024, Bunney2025, Torkzadeh2026}
as a reference: three of the four adatom valence electrons go into the three covalent bonds to the substrate, and the fourth half-fills the apical $p_z$-derived ($a_1$) band.

On the polar cBN(111) surfaces, in contrast, this counting is modified by the electron surplus or deficit required to saturate the substrate bonds. B contributes $3/4=0.75$ electrons per bond, N $5/4=1.25$, compared to one electron per bond in the group-IV case. The adatom must compensate this deficit or surplus to complete the three covalent bonds, supplying 3.75 (2.25) electrons in total on the B- (N-)termination and leaving 0.25 (1.75) electrons in the $p_z$-derived band. Both fillings lie close to the empty and full limits, where correlation effects are weak, so we do not consider group-IV adatoms on cBN(111) further.

Group-V and group-III adatoms shift these fillings into the correlated regime. Relative to the group-IV@B and group-IV@N cases, their one additional or one fewer valence electron, respectively, moves the $p_z$ band filling by exactly one electron: 1.25 electrons for group-V@B, and 0.75 electrons for group-III@N.

The filling of the two-band systems formed by group-III@B follows the same logic, starting from the hypothetical group-IV@B reference. Removing the one valence electron that distinguishes group-III from group-IV first empties the $p_z$ band, and the remaining 0.75 electrons must then be removed from the energetically lower, otherwise fully occupied planar $e$ doublet, so that $E_\mathrm{F}$ moves into this doublet, which retains $4-0.75=3.25$ electrons.

The TM systems can be analyzed analogously to their counterparts on 3C-SiC(111)~\cite{menke2025}, where three covalent bonds to the group-IV substrate absorb three valence electrons and the crystal field places the $e$ doublet at $E_\mathrm{F}$, leaving one, two, and three electrons in it for Ti, V, and Cr, respectively.
The crystal field on B-terminated cBN(111) is analogous, but saturating the three bonds consumes 3.75 electrons, leaving 1.25, 2.25, and 3.25 electrons in the $e$ doublet for V, Cr, and Mn, respectively.

Since the bottom surface is passivated (Sec.~\ref{section:method}), it hosts no states in the band gap and exchanges no charge with the adatom layer. The non-integer fillings are, hence, fixed by the bonding chemistry of the polar cBN(111) surface alone.
Note that, in contrast to a semiconductor, where such self-doping merely shifts $E_\mathrm{F}$, a quarter of an electron per site triggers the Mott insulator-to-metal phase transition (see Sec.~\ref{section:dmft}).
\makeatletter\@booleantrue\force@deferlist@sw\makeatother
\subsection{cRPA}
Table~\ref{tab:overview} lists the Hubbard-Kanamori parameters obtained from cRPA for the correlated subspace of each system: the Hubbard $U$ for the one-band systems, and $U$, $U'$ and $J_\mathrm{H}$ for the two-band systems. For the latter, the symmetry of the $e$ doublet enforces $U'=U-2J_\mathrm{H}$, which the (unrounded) cRPA values fulfill to within \SI{1}{\milli\eV}.

For the one-band systems, $U$ decreases with increasing period of the adatom, from \SI{4.35}{\eV} (N) to \SI{2.83}{\eV} (As) on the B-termination and from \SI{4.23}{\eV} (B) to \SI{1.92}{\eV} (Ga) on the N-termination, while the Wannier spread grows (Table~\ref{tab:overview}). As for group-IV adatoms on Si(111)~\cite{hansmann2013jpcm}, this trend reflects the larger extent of the valence $p$ orbitals of heavier elements. The group-III adatoms on the B-termination follow the same pattern: the light B adatom has the largest interaction, $U=\SI{3.20}{\eV}$, while Al and Ga are almost indistinguishable, with $U=\SI{2.56}{\eV}$ and \SI{2.58}{\eV} (consistent with their nearly identical band dispersions).

Across the TM series, $U$ increases from \SI{1.33}{\eV} (V) to \SI{1.79}{\eV} (Mn) as the $3d$ Wannier orbitals become more localized, analogous to the TM adatoms on 3C-SiC(111)~\cite{menke2025}. At comparable spatial extent, however, the TM systems have a much smaller $U$ than the $p$-orbital systems, e.g., \SI{1.79}{\eV} for Mn@B compared with \SI{4.35}{\eV} for N@B.

The role of the substrate becomes apparent when the cRPA interaction is separated into the unscreened interaction $U_\mathrm{bare}$, which measures the localization of the Wannier orbital, and the screening ratio $U/U_\mathrm{bare}$ (Table~\ref{tab:screening}).
For the $p$-orbital systems, both quantities exceed those of the group-IV adatoms on Si(111)~\cite{hansmann2013jpcm}. The large interaction of these systems thus results from the stronger localization of the adatom orbitals as well as from the weaker dielectric screening of cBN.
For the TM systems, in contrast, the screening is dominated by the remaining adatom $3d$ states above the correlated $e$ doublet (Fig.~\ref{fig:dft_all}). As for the TM adatoms on SiC~\cite{menke2025}, the strongly dispersive $a_1$-derived band lies closest to the doublet and thus contributes significantly to the screening. These states reduce the bare interaction to about $11\%$ on both substrates. Being largely internal to the adatom, this screening hardly depends on the substrate, and the larger $U$ on cBN than on SiC stems mainly from the stronger localization of the $3d$ orbitals, reflected in the larger $U_\mathrm{bare}$.

We computed the nearest-neighbor interaction $V_{nn}$, i.e., the screened interaction between Wannier orbitals on neighboring adatoms, for V@B, Cr@B, and As@B (Supplemental Material~\cite{Supplemental}). With $V_{nn}/U=0.37$--$0.42$, it is sizable and comparable to prior cRPA results for adatoms on Si(111)~\cite{hansmann2013jpcm} and 3C-SiC(111)~\cite{menke2025}. For C:Si(111), whose interaction parameters are similar, nearest-neighbor and longer-range interactions were shown to have negligible effects beyond the mean-field level~\cite{hansmann2013prl}. We therefore consider only local interactions in DMFT, as in Ref.~\cite{menke2025}.
\begin{table}[tb]
\centering
\caption{Ranges of the unscreened ($U_\mathrm{bare}$) and screened interaction $U$, and their ratio $U/U_\mathrm{bare}$ for the adatom systems on cBN(111), compared with group-IV adatoms on Si(111)~\cite{hansmann2013jpcm} and TM adatoms on 3C-SiC(111)~\cite{menke2025}.
}
\begin{tabular*}{\linewidth}{
    @{} l @{\extracolsep{\fill}}
    r @{\extracolsep{0pt}--} l @{\extracolsep{\fill}}
    r @{\extracolsep{0pt}--} l @{\extracolsep{\fill}}
    r @{\extracolsep{0pt}--} l @{}
}
\toprule
adatoms & \multicolumn{2}{c}{$U_\mathrm{bare}$ [\si{\eV}]} & \multicolumn{2}{c}{$U$ [\si{\eV}]} & \multicolumn{2}{c}{$U/U_\mathrm{bare}$} \\
\midrule
group IV on Si(111) ($p$) & 4.3  & 6.0  & 0.9 & 1.4 & 0.21 & 0.23 \\
group III/V on cBN ($p$)  & 6.1  & 12.3 & 1.9 & 4.4 & 0.31 & 0.41 \\
\midrule
TM on 3C-SiC ($d$)        & 7.2  & 12.8 & 0.9 & 1.4 & 0.11 & 0.12 \\
TM on cBN ($d$)           & 11.2 & 16.1 & 1.3 & 1.8 & 0.11 & 0.12 \\
\bottomrule
\end{tabular*}
\label{tab:screening}
\end{table}

\begin{figure*}[t]
\includegraphics[width=2\columnwidth]{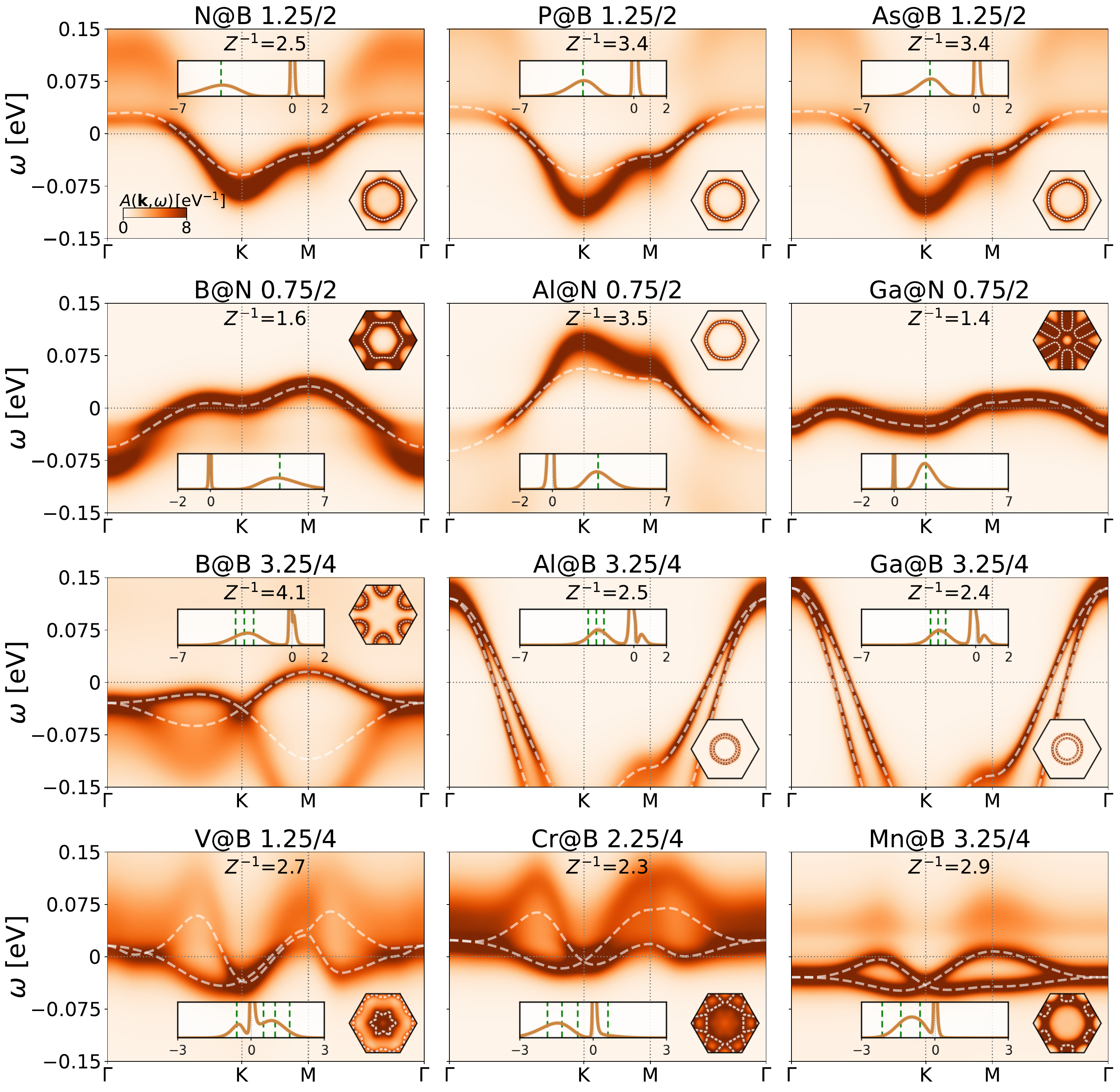}
\caption{DMFT spectral functions $A(\mathbf{k},\omega)$ (zoomed in around $E_\mathrm{F}$) from charge-self-consistent calculations at $T\approx\SI{116}{\kelvin}$. The color scale gives $A(\mathbf{k},\omega)$ in \si{\per\eV}, and the width of the quasiparticle band reflects the broadening $-\imag\Sigma(\omega)$ shown in Fig.~\ref{Fig-sigma}. White dashed lines show the renormalized non-interacting bands $Z\,\varepsilon(\mathbf{k})$, with the corresponding $m^*/m=Z^{-1}$ given in each panel. Rows as in Fig.~\ref{fig:dft_all}. The rectangular insets show the local ($k$-integrated) spectral function (DOS) over a wide energy window (axis in eV), highlighting the position of the Hubbard band relative to the quasiparticle peak at $E_\mathrm{F}$. Solid lines denote the average and shaded envelopes the analytic-continuation error bars. Green dashed lines mark the atomic excitation energies, $-U$ ($+U$) for the one-band systems on the B- (N-)termination and the multiplet lines (Supplemental Material~\cite{Supplemental})
for the two-band systems. The hexagonal insets show the DMFT and quasiparticle Fermi surfaces (white dashed lines).
}
\label{fig:dmft_spectral}
\end{figure*}
\begin{figure}[t]
\includegraphics[width=\columnwidth]{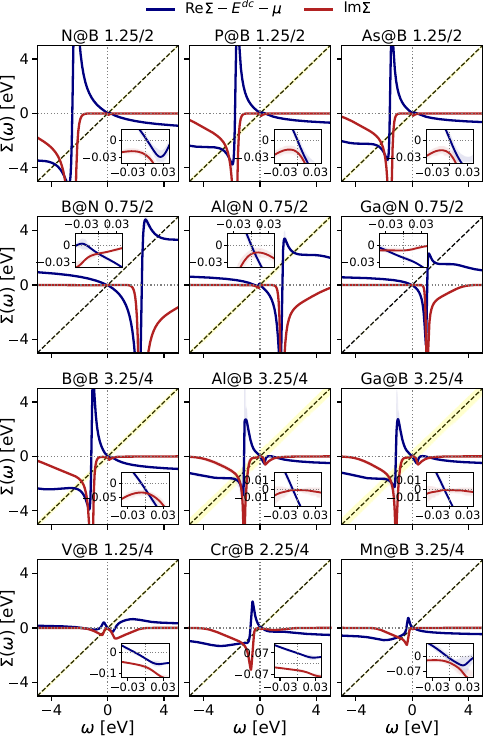}
\caption{DMFT self-energies at $T\approx\SI{116}{\kelvin}$: $\real\Sigma(\omega)-E^\mathrm{dc}-\mu$ (blue), with the double-counting correction $E^\mathrm{dc}$ and the chemical potential $\mu$, and $\imag\Sigma(\omega)$ (red). Solid lines denote the average and the blue and red shaded envelopes the analytic-continuation error bars. The dashed line is the angle bisector, the yellow shaded area the width of the non-interacting bands. The insets zoom into $|\omega|\leq\SI{0.04}{\eV}$ around $E_\mathrm{F}$ (individual vertical scales). Rows as in Fig.~\ref{fig:dft_all}.
}
\label{Fig-sigma}
\end{figure}

\section{Results: DMFT}
\label{section:dmft}
In Fig.~\ref{fig:dmft_spectral} we show the momentum-resolved spectral functions $A(\mathbf{k},\omega)$ of all twelve systems near $E_\mathrm{F}$. In Fig.~\ref{Fig-sigma} we show the corresponding self-energies.
In all twelve systems, the charge self-consistency leaves the adatom-derived Wannier bands practically unchanged.
The low-energy physics is hence well described by the Hubbard models of Sec.~\ref{section:method} (hoppings in the Supplemental Material~\cite{Supplemental}, interaction parameters in Table~\ref{tab:overview}), and these models can be used as they are in further studies.
The ratio $U/W$ between the Hubbard interaction and the bandwidth ranges from $2.6$-$2.7$ (Ga@B, Al@B) to $30$-$38$ (B@N, Ga@N). Despite these large ratios, we find no Mott insulator. At the inverse temperature $\beta=\SI{100}{\per\eV}$ ($T\approx\SI{116}{\kelvin}$) considered here, all twelve systems remain ungapped and exhibit spectral weight at $E_\mathrm{F}$.
This is a consequence of the self-doping, which places the filling $0.25$ electrons away from an integer reference filling $n_\mathrm{ref}$ in every case. At $n_\mathrm{ref}$, the same parameters would be expected to yield a Mott insulator for most systems, as realized for the analogous V and Cr adatoms on 3C-SiC(111)~\cite{menke2025} and shown for Cr@B at the end of this section.
The spectra follow from the self-energy as the imaginary part of the interacting retarded lattice Green's function,
\begin{align}
A(\mathbf{k},\omega)&=-\tfrac{1}{\pi}\,\imag G(\mathbf{k},\omega), \label{eq:spectral}\\
G(\mathbf{k},\omega)&=\frac{1}{\omega+i0^{+}+\mu-\varepsilon(\mathbf{k})-\Sigma(\omega)}, \label{eq:green}
\end{align}
with $\mu$ the chemical potential and $\Sigma(\omega)$ the retarded self-energy, which by symmetry is the same for both orbitals of the two-band systems. $\real\Sigma(\omega)$ acts as an energy-dependent shift of the non-interacting levels $\varepsilon(\mathbf{k})$, and $-\imag\Sigma(\omega)$ as a broadening.
From the low-energy slope of $\real\Sigma(\omega)$ we extract effective mass renormalizations
$m^*/m=Z^{-1}$ with $Z=[1-\partial_\omega\real\Sigma(\omega)|_{\omega=0}]^{-1}$, ranging from $1.4$ (Ga@N) to $4.1$ (B@B) (Table~\ref{tab:overview}).
Due to the large $U/W$ ratios, only Al@B and Ga@B reach a Fermi-liquid regime at $T\approx\SI{116}{\kelvin}$. For the other systems, the analytically continued $\real\Sigma(\omega)$ shows no extended linear regime around $\omega=0$ (insets in Fig.~\ref{Fig-sigma}).
The resulting $Z$ should therefore not be read as a quasiparticle (QP) weight, which is well defined only in the Fermi-liquid regime. Rather, it is an effective renormalization factor for the DFT bands that reproduces the QP dispersion (and Fermi surface) near $E_\mathrm{F}$ (white dashed lines in Fig.~\ref{fig:dmft_spectral}). Dispersive QP-like excitations thus persist above the Fermi-liquid coherence temperature in line with the ``resilient quasiparticles'' found in doped Mott insulators~\cite{Deng2013}.

\subsection{One-band systems}
All one-band systems exhibit a single broad satellite at $\omega\approx\pm U$, whose side is determined by the doping direction: it lies below $E_\mathrm{F}$ for group-V@B ($1.25/2$ electrons) and above $E_\mathrm{F}$ for group-III@N ($0.75/2$ electrons).
This is the behavior of a doped Mott insulator in the atomic limit ($U\gg W$). Electron/hole doping pins $E_\mathrm{F}$ to the upper/lower Hubbard band, which then develops into the QP peak, while the opposite Hubbard band survives as an incoherent satellite at a distance $U$ from $E_\mathrm{F}$, i.e., the full atomic charge gap.
To make this explicit, $\pm U$ is marked by green dashed vertical lines in the local spectral functions (insets) of the one-band systems in Fig.~\ref{fig:dmft_spectral}. The satellite maxima coincide with $\pm U$ to within $5\%$ (Table~\ref{tab:overview}), although $U$ varies by a factor of $2.3$ and $W$ by almost an order of magnitude across the six systems.

The self-energy of each one-band system (Fig.~\ref{Fig-sigma}) is dominated by a single sharp resonance on the satellite side, at $|\omega_\mathrm{res}|=(0.54$--$0.57)\,U$, i.e., between the QP peak and the satellite at $\pm U$.
This distinguishes a Hubbard band of atomic origin from the incoherent weight of a moderately correlated Fermi liquid with $Z<1$~\cite{Meinders1993,Rozenberg1996}:
a smooth self-energy transfers weight $1-Z$ away from the QP peak without generating a separate feature, whereas a pole-like resonance depletes the spectral weight at its position and thereby splits off a separate band.

In the atomic limit, the band around $E_\mathrm{F}$ carries the spectral weight $w=(1+\delta)/2$ per spin at the filling $n=1\pm\delta$.
Within the Hubbard-I approximation~\cite{Hubbard1963}, which treats the hopping as a perturbation of the atom, this band follows the noninteracting dispersion scaled by the same factor, $\varepsilon_{\mathbf{k}}\to w\,\varepsilon_{\mathbf{k}}$, so that the mass renormalization approaches the universal value
\begin{equation}
Z^{-1}\;\to\;\frac{1}{w}=\frac{2}{1+\delta}=1.60
\end{equation}
at $\delta=0.25$ (see Supplemental Material~\cite{Supplemental}).

Note that, in contrast to a Mott transition, which requires integer filling, $Z^{-1}$ thus stays finite for $U/W\to\infty$.
The two systems with the largest $U/W$, Ga@N and B@N ($38$ and $30$), come closest to this limit, with $Z^{-1}=1.4$ and $1.6$, whereas the other one-band systems lie at $2.5$-$3.5$ (Table~\ref{tab:overview}).
Hubbard-I neglects the coherence of the QPs and is exact only at temperatures well above the coherence scale, which is proportional to the bandwidth. A large $U/W$ alone is therefore not sufficient, and indeed $Z^{-1}$ increases with $W$, from Ga@N ($W=\SI{0.05}{\eV}$) over B@N and N@B to Al@N ($W=\SI{0.41}{\eV}$).
The self-energy pole, in contrast, is set by the atomic energy scale: Hubbard-I places it at $|\omega_\mathrm{res}|=w\,U=0.625\,U$, close to the observed values irrespective of $U/W$.\\

\textit{Quasiparticles:}
At $E_\mathrm{F}$ we observe sharp quasiparticle features in the DMFT spectral function.
The corresponding self-energies (Fig.~\ref{Fig-sigma}), however, are not of Fermi-liquid form: near $\omega=0$, $-\mathrm{Im}\,\Sigma(\omega)$ does not follow the quadratic $\omega^2$ dependence and retains a finite scattering rate at $E_\mathrm{F}$. This indicates that $T\approx\SI{116}{\kelvin}$ lies above the Fermi-liquid coherence temperature.
Nonetheless, the DMFT Fermi surfaces closely follow the DFT contours for all group-V@B and Al@N\footnote{at low enough $T$ they must coincide: with a $\mathbf{k}$-independent DMFT self-energy, the Fermi surface is the contour $\varepsilon(\mathbf{k})=\mu-\real\Sigma(0)$, whose enclosed area is fixed by the filling through Luttinger's theorem~\cite{Luttinger1960,MuellerHartmann1989}}.
In contrast, B@N and Ga@N have extremely flat dispersions which, together with temperature and lifetime broadening, lead to a spread of the spectral weight over extended regions of the Brillouin zone, which makes the identification of a Fermi surface rather questionable.

The coherence of the QPs can be quantified by the scattering rate at $E_\mathrm{F}$ relative to the bandwidth, $-\imag\Sigma(0)/W$ (Table~\ref{tab:overview}), which equals the ratio of the QP damping $-Z\imag\Sigma(0)$ to the QP bandwidth $ZW$.
Ga@N has the smallest absolute scattering rate of all one-band systems ($\approx\SI{9}{\milli\eV}$), negligible on the scale of Fig.~\ref{Fig-sigma}, but relative to its narrow band it is the least coherent ($-\imag\Sigma(0)/W=0.18$). Al@N has the widest band and is the most coherent ($0.04$).\\

\begin{figure}[t]
\includegraphics[width=\columnwidth]{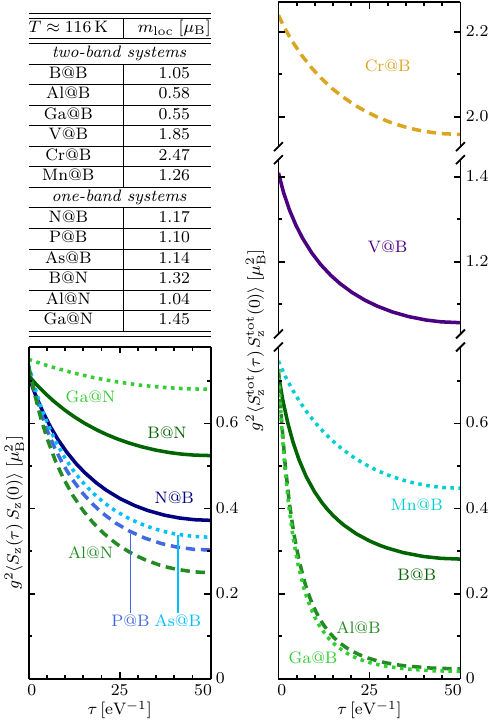}
\caption{Local spin susceptibilities $\chi_\text{loc}(\tau)$: Left for one-band systems and on the right for two-band systems, where the operator $S^\text{tot}_z$ denotes the $z$-component of the spin summed over both orbitals. Table of local magnetic moments for all systems (top left).}
\label{Fig-Sus}
\end{figure}

\textit{Local moments:} The renormalized QPs coexist with localized magnetic moments which we can estimate through the (dynamic) local spin susceptibility following~\cite{menke2025, Hansmann2010, Toschi2012}:
\begin{equation}
\label{eq:chi}
    \chi_\text{loc}(\tau)
    = g^2\mu_B^2 \sum_{i,j} \langle \mathcal{T}_\tau\,
        S_z^i(\tau)\, S_z^{j}(0) \rangle,
\end{equation}
with $S_z^i(\tau) = \tfrac{1}{2}\big[n^i_\uparrow(\tau) - n^i_\downarrow(\tau)\big]$ the $z$-component of the spin operator of orbital $i$ and $g$ the $g$-factor.
From the Fourier transform of Eq.~\eqref{eq:chi} at $\omega=0$ we obtain the local moment as $m_\text{loc} \simeq \sqrt{3\,k_\text{B}T\, \chi_\text{loc}(\omega=0)}$.
In Fig.~\ref{Fig-Sus} we plot the dynamic spin susceptibility $\chi_\text{loc}(\tau)$ for all systems and list their local magnetic moments.

$\chi_\text{loc}(\tau)$ characterizes the interplay of localized moments and renormalized QPs on different timescales~\cite{Watzenboeck2020}. In all systems, it decays from $\chi_\text{loc}(\tau=0)$ to $\chi_\text{loc}(\tau=\beta/2)$ due to metallic screening and in contrast to the near-constant $\chi_\text{loc}(\tau)$ of the Mott-insulating V and Cr adatoms on 3C-SiC(111) at the same temperature~\cite{menke2025}.
For the six one-band systems, the instantaneous response $\chi_\text{loc}(\tau=0)=0.71$--$0.75~\mu_B^2$ is close to the atomic value $0.75~\mu_B^2$ for a deviation of $\pm0.25$ electrons from half filling, whereas the screening grows with the QP bandwidth: $m_\text{loc}$ decreases monotonically from $1.45~\mu_B$ for Ga@N ($ZW\approx\SI{0.04}{\eV}$) to $1.04~\mu_B$ for Al@N ($ZW\approx\SI{0.12}{\eV}$).

\subsection{Two-band systems}
All two-band systems are electron doped with respect to the integer reference filling, $n=n_\mathrm{ref}+\delta$ with $\delta=0.25$, so that $E_\mathrm{F}$ is pinned to the upper Hubbard band, which develops into the QP peak. In contrast to the one-band systems, the Hubbard bands are not single atomic lines: Hund's coupling splits the two-electron configuration into a high-spin triplet and singlets $2J_\mathrm{H}$ and $4J_\mathrm{H}$ above it, so that each satellite comprises several multiplet lines (green dashed lines in the insets of Fig.~\ref{fig:dmft_spectral}). These lines are not resolved individually, but the mean energies of all satellites agree with the weighted means of the corresponding atomic lines to within \SI{0.1}{\eV} (Supplemental Material~\cite{Supplemental}). The distance of the satellite below $E_\mathrm{F}$ is therefore set by $J_\mathrm{H}$ through the reference configuration, $U-3J_\mathrm{H}$ for $n_\mathrm{ref}=1$, $U+J_\mathrm{H}$ (dominant line) for $n_\mathrm{ref}=2$, and on average $U-\tfrac53J_\mathrm{H}$ for $n_\mathrm{ref}=3$, and ranges from $0.48\,U$ (V@B) to $1.06\,U$ (Cr@B). Only the Fermi liquids Al/Ga@B deviate from the atomic picture: their satellite below $E_\mathrm{F}$ retains only $\approx80\%$ of the atomic weight ($\gtrsim90\%$ in the other two-band systems), and a self-energy resonance splits off a third peak at $\approx\SI{0.5}{\eV}$ that takes weight from the QP peak and has no atomic counterpart. \\

\textit{Quasiparticles:}
For group-III@B, the DMFT Fermi surfaces (Fig.~\ref{fig:dmft_spectral}) remain qualitatively identical to their DFT counterparts. For TMs, flat-band behavior close to $E_\mathrm{F}$ spreads the spectral weight over extended regions of the Brillouin zone.
The renormalized bands $Z\varepsilon(\mathbf{k})$ track the QP dispersion over most of the bandwidth only for the Fermi liquids Al/Ga@B; elsewhere, the energy dependence of $\real\Sigma(\omega)$ causes deviations that grow with distance from $E_\mathrm{F}$.
The Dirac crossings of the QP bands at $K$ lie close to $E_\mathrm{F}$ (\SI{10}{\milli\eV} below for Cr@B, $\approx\SI{40}{\milli\eV}$ for B/Mn/V@B), except in Al/Ga@B ($\approx\SI{0.25}{\eV}$).\\
\begin{figure*}[t]
\includegraphics[width=2\columnwidth]{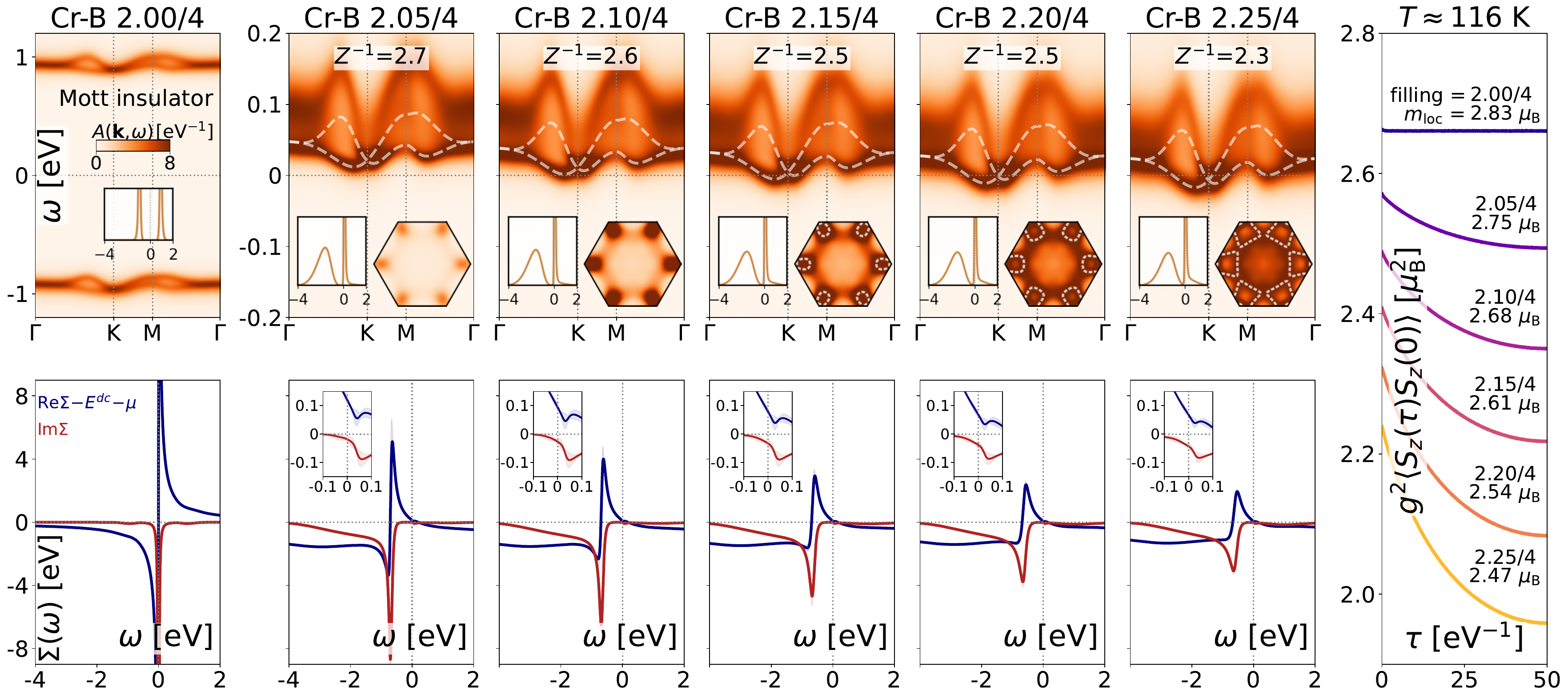}
\caption{Cr@B at fillings $n=2.00$ to $2.25$. Top row: momentum-resolved spectral function $A(\mathbf{k},\omega)$.
White dashed lines show the renormalized non-interacting bands $Z\,\varepsilon(\mathbf{k})$, with the corresponding $m^*/m=Z^{-1}$ given in each panel. Insets show the local spectral function $A(\omega)$ and the DMFT Fermi surfaces, with the quasiparticle Fermi surfaces as white dashed lines. At $n=2.00$ (Mott insulator) there is no Fermi surface and no quasiparticle band. Bottom row: real-frequency self-energy.
Right: local spin susceptibility $\chi_\text{loc}(\tau)$, labeled by filling and local moment.}
\label{fig:cr_doped}
\end{figure*}

\textit{Local moments:}
All two-band systems at $3.25/4$ filling (B/Al/Ga/Mn@B) show the same instantaneous response as the one-band systems, $\chi_\text{loc}(\tau=0)=0.71$--$0.75~\mu_B^2$, close to the atomic value $0.75~\mu_B^2$ of a $3{:}1$ mixture of $S=\tfrac12$ and $S=0$ sites.
Their screening follows the QP bandwidth $ZW$, from Al/Ga@B, where $\chi_\text{loc}(\tau=\beta/2)$ nearly vanishes (as expected for Fermi liquids well below the coherence temperature) and the moments are smallest, up to Mn@B, see Table~\ref{tab:overview} and Fig.~\ref{Fig-Sus}.
The largest moments occur for Cr@B ($2.47~\mu_B$) and V@B ($1.85~\mu_B$), where a fraction $1-\delta$ and $\delta$ of the sites, respectively, carries two electrons.
In the atomic limit, Hund's coupling binds these into a high-spin triplet: the inter-orbital correlation $2\langle S_z^aS_z^b\rangle=\tfrac16$ raises $\langle(S_z^a+S_z^b)^2\rangle$ from $\tfrac12$ for uncorrelated spins to $\tfrac23$, so that an $S=1$ site responds with $\tfrac83~\mu_B^2$, $\tfrac43$ times more than without this correlation.
Accordingly, $\chi_\text{loc}(\tau=0)=2.24$ (Cr@B) and $1.41~\mu_B^2$ (V@B) match the atomic mixtures of $S=1$ and $S=\tfrac12$ sites, $(1-\delta)\tfrac83+\delta=2.25$ and $(1-\delta)+\tfrac83\delta=1.42~\mu_B^2$.
Despite larger QP bandwidths ($ZW\approx86$ and \SI{106}{\milli\eV} vs.\ \SI{56}{\milli\eV} for Mn@B), these moments are screened much less ($\chi_\text{loc}(\beta/2)/\chi_\text{loc}(0)=0.87$ and $0.75$ vs.\ $0.60$), consistent with the reduced coherence scale of Hund's-coupled moments~\cite{georges2013strong}.

\subsection{Overview}
\begin{enumerate}

\item Among the one-band systems, P@B, As@B and Al@N ($U/W=7$--$9$) have the most coherent QPs and strongly screened moments. They realize the doped triangular-lattice Hubbard model at $\delta=\pm\sfrac14$ and are natural candidates to test whether the unconventional superconductivity predicted at weaker coupling~\cite{cao2018,huang2022topological,huang2023chiral} persists in this regime.
\item The narrow-band systems B@N and Ga@N ($U/W=30$ and $38$) carry the largest, least screened moments of the one-band systems and have the least coherent QPs. They are candidates for frustrated magnetism at lower temperatures.
\item Among the two-band systems, Al@B and Ga@B are only moderately correlated, whereas Cr@B is the most promising: it has a half-filled $S=1$ reference, is closest to the atomic limit ($\Delta_\mathrm{at}/W=9.4$\footnote{With two bands, Hund's coupling changes the atomic charge gap $\Delta_\mathrm{at}$, the cost of moving an electron between two atoms at $n_\mathrm{ref}$, from $U$ to $U+J_\mathrm{H}$ at half filling and to $U-3J_\mathrm{H}$ for $n_\mathrm{ref}=1,3$~\cite{demedici2011janus}.}), carries the largest local moment ($2.47~\mu_B$), and has its Dirac crossing only \SI{10}{\milli\eV} below $E_\mathrm{F}$.
\end{enumerate}
\subsection{External doping in Cr@B}
\label{sec:cr_doped}
Cr@B is the natural candidate to isolate the effect of the substrate-imposed non-integer filling, as its half-filled $S=1$ reference matches that of V adatoms on 3C-SiC(111), which form a robust Mott insulator~\cite{menke2025}.
Fig.~\ref{fig:cr_doped} shows spectra, self-energies and $\chi_\text{loc}(\tau)$ for fixed hopping and interaction parameters, with $n$ varied from $2$ to the actual $2.25$ in steps of $0.05$.
At $n=2$, Cr@B is a Mott insulator: the Hubbard bands are centered at $\pm(U+J_{\mathrm{H}})/2=\pm\SI{0.94}{\eV}$, the self-energy has a pole at $\omega=0$, and $\chi_\text{loc}(\tau)$ is constant at the unscreened values of an atomic $S=1$ site ($m_\mathrm{loc}=2.83~\mu_B$).
Already $\delta=0.05$ establishes the spectral structure found at $\delta=0.25$: $E_\mathrm{F}$ moves into the former upper Hubbard band, which becomes the QP band, and the lower Hubbard band remains as the only satellite, $\approx U+J_{\mathrm{H}}$ below $E_\mathrm{F}$.
With increasing $\delta$, the screening grows ($\chi_\text{loc}(\beta/2)/\chi_\text{loc}(0)=0.97\to0.87$) and $m_\mathrm{loc}$ drops from $2.75$ to $2.47~\mu_B$, whereas $\chi_\text{loc}(\tau=0)$ follows the atomic mixture of $S=1$ and $S=\tfrac12$ sites, $[(1-\delta)\,\tfrac83+\delta]~\mu_B^2$, within $0.5\%$ at every filling.
The mean energy and weight of the satellite agree with the atomic values within $2\%$ at every filling (Supplemental Material~\cite{Supplemental}).
Cr@B is thus a doped Mott insulator in the literal sense: the self-doping of cBN(111) turns the upper Hubbard band into the QP band, while the lower one and the instantaneous spin response remain atomic-like.
\section{Conclusions}
\label{section:conclusions}
We have investigated twelve adatom lattices on the B- and N-terminated cBN(111) surface using DFT, cRPA, and fully charge self-consistent DMFT: group-V@B and group-III@N realize one-band Hubbard models, whereas group-III@B and TM@B (V, Cr, Mn) yield two-band Hubbard models.
The impact of charge self-consistency is negligible, so the low-energy physics of all systems is well described by these Hubbard models, with ratios $U/W=2.6$-$38$.
The polarity of cBN(111) self-dopes the lattices on the B/N-termination by $\sfrac14$ electron/hole per adatom away from integer filling.

At $T\approx\SI{116}{\kelvin}$ we find all systems to be correlated metals, which is a consequence of the modulation doping by the cBN(111) substrate.
External doping can, hence, act opposite to the usual scenario of a Mott insulator that is doped into a metal: In Cr@B, the removal of $0.25$ electrons per adatom dopes the lattice back into the $S=1$ Mott insulator at integer filling.

Whether the unconventional superconductivity predicted for doped Mott insulators on the triangular lattice persists at the stronger coupling realized here is an open question.
Their large local moments (up to $2.5~\mu_B$ for Cr@B) make most of them also candidates for frustrated magnetism at lower temperatures, as reported for Sn/Si(111)~\cite{gangli2013}.
We propose cBN(111) as a clean platform for this physics.

\begin{acknowledgments}
We thank Ralph Claessen, Jörg Schäfer, Simon Moser, and Giorgio Sangiovanni for fruitful discussions. The authors gratefully acknowledge the scientific support and HPC resources provided by the Erlangen National High Performance Computing Center (NHR@FAU) of the Friedrich-Alexander-Universit\"at Erlangen-N\"urnberg (FAU) under the NHR project b181dc. NHR@FAU hardware is partially funded by the German Research Foundation (DFG) -- 440719683. N.E. and J.M. acknowledge partial funding by the DFG Research Training Group GRK2861 -- 491865171. 
\end{acknowledgments}

\bibliography{refs}
\clearpage
\appendix
\section{Stability of slab geometries}
\begin{table}[h]
  \centering
  \caption{Total-energy difference $\Delta E_\mathrm{tot}=E_\mathrm{tot}(\mathrm{T}_4)-E_\mathrm{tot}(\mathrm{H}_3)$
  between the relaxed T$_4$ and H$_3$ adsorption geometries, per $(\sqrt{3}\times\sqrt{3})R30^\circ$ cell, i.e., per
  adatom. Negative (positive) values favor the T$_4$ (H$_3$) site. In this work, the H$_3$ site is used for Al@N and
  Ga@N, and the T$_4$ site for all other systems.}
  \label{tab:adsorption_energy_difference}
  \begin{tabular}{l S[table-format=+1.2, retain-explicit-plus]}
  \toprule
  system & {$\Delta E_\mathrm{tot}$ [\si{\eV}]} \\
  \midrule
  N@B  & -0.58 \\
  P@B  & -0.64 \\
  As@B & -0.52 \\
  \hline
  B@N  & -0.46 \\
  Al@N & +0.11 \\
  Ga@N & +0.00 \\
  \hline
  B@B  & -1.48 \\
  Al@B & -0.44 \\
  Ga@B & -0.33 \\
  \hline
  V@B  & -0.65 \\
  Cr@B & -0.89 \\
  Mn@B & -1.05 \\
  \bottomrule
  \end{tabular}
\end{table}
Table~\ref{tab:adsorption_energy_difference} compares the total energies of the T$_4$ and H$_3$ adsorption geometries, obtained from the structural optimizations described in the main text for slabs with eight BN bilayers.

\begin{figure*}[t]
\includegraphics[width=2\columnwidth]{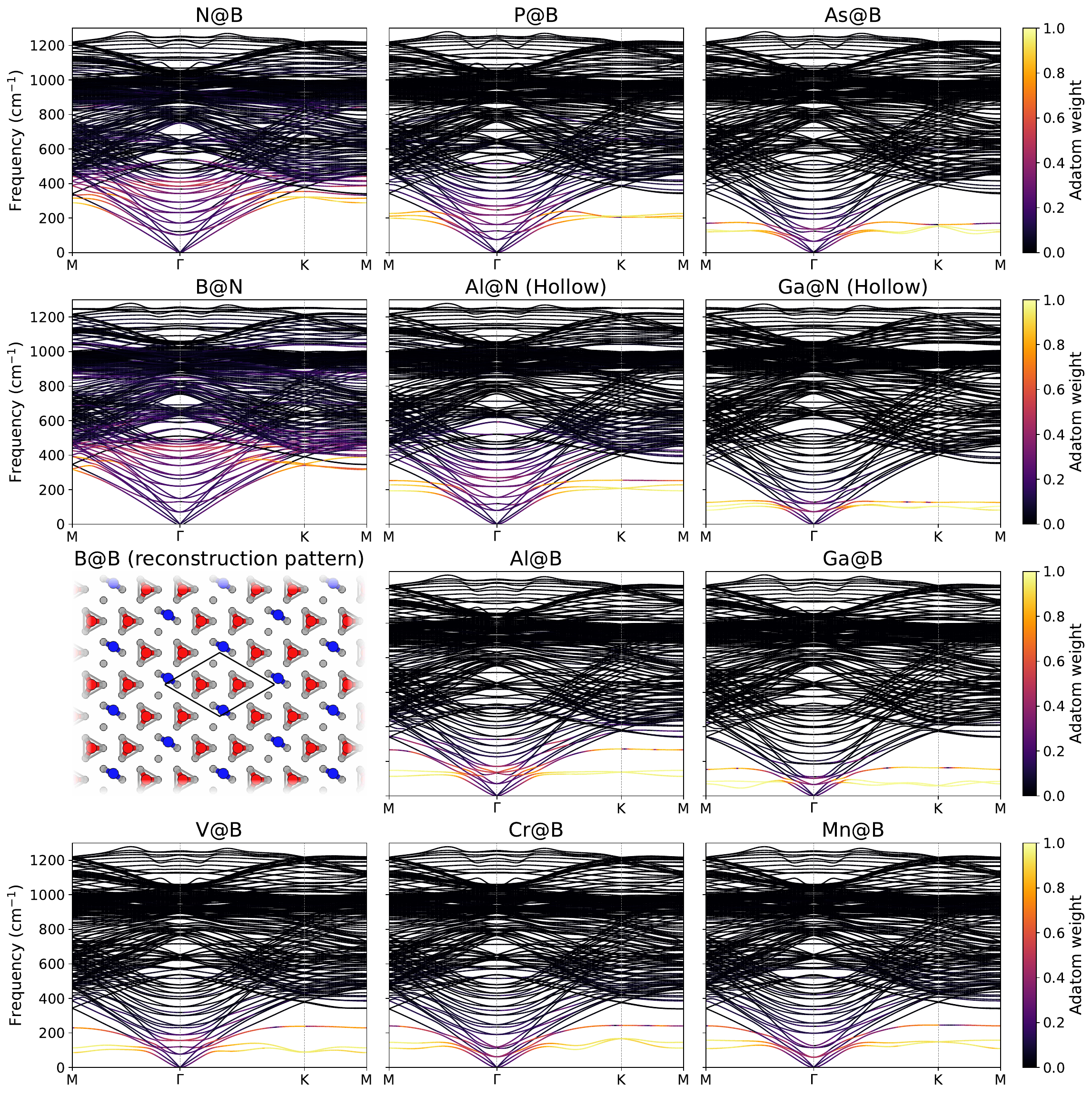}
\caption{Phonon dispersion of cBN(111) surface slabs (10 BN bilayers) with adatoms at 1/3-ML coverage along
the high-symmetry path of the hexagonal surface Brillouin zone. Rows correspond to: (1) group-V adatoms (N, P, As) on the boron surface; (2) group-III adatoms (B, Al, Ga) on the nitrogen surface; (3) group-III adatoms (B, Al, Ga) on the boron surface; (4) $3d$ TM adatoms (V, Cr, Mn) on the boron surface. The color scale depicts the contribution of the adatoms to the phonon modes. For B@B, whose high-symmetry geometry is dynamically unstable at the semi-local (PBE) level, the panel instead shows the assumed reconstruction: the enlarged unit cell (black lines) contains two adatoms at T$_4$ sites (red) and one displaced toward an H$_3$ site (blue).}
\label{phonons}
\end{figure*}

The dynamical stability of the resulting adatom lattices was assessed from phonon dispersions computed within density functional perturbation theory (DFPT)~\cite{baroni2001phonons} as implemented in Quantum ESPRESSO~\cite{giannozzi2009quantum,giannozzi2017advanced}.
Following the phonon calculations for TM adatoms on SiC in Ref.~\cite{menke2025}, a vacuum region of \SI{30}{\angstrom} was used and the slab thickness was increased to ten BN bilayers (passivated by pseudo-hydrogen
as in the main text).
The in-plane lattice constant was fixed to the DFT-optimized bulk value, so that all adatom lattices share the periodicity of the substrate, as they would on a cBN(111) surface in experiment.
We used norm-conserving PBE pseudopotentials from the PseudoDojo library~\cite{van2018pseudodojo}, a wavefunction cutoff of \SI{120}{\Ry}, a Gaussian smearing of \SI{0.015}{\Ry} and a $12\times12\times1$ $\mathbf{k}$-point mesh. All atomic positions were relaxed until the residual forces fell below \SI[per-mode=symbol]{e-3}{\eV\per\angstrom}.
To rule out a bias from the symmetric starting configuration, relaxations were also started from displaced adatom positions, and in all cases the adatoms returned to the high-symmetry sites.
Dynamical matrices were computed on a $6\times6\times1$ $\mathbf{q}$-point mesh and transformed to real-space force constants. On these we imposed the acoustic sum rules for translational and rotational invariance and the 15 Huang conditions for a stress-free equilibrium~\cite{Lin2022_phonons_asr}. Fourier interpolation then yielded the dispersions shown in Fig.~\ref{phonons} along the high-symmetry path of the surface Brillouin zone.

All adatom lattices are free of imaginary phonon modes and -- within DFT -- dynamically stable. B@B is the sole exception. It exhibits a doubly degenerate unstable mode at $K$ with a frequency of \SI{-170}{\per\centi\meter} (negative values denote imaginary frequencies).
We displaced the B adatoms along the eigenvector of this mode and relaxed the structure in a $3\times3$ supercell containing nine adatoms, which is commensurate with the $K$ point. To keep these calculations tractable, the slab thickness was reduced to three BN bilayers.
The relaxation converged to a reconstructed structure with a $\sqrt{3}\times\sqrt{3}$ periodicity relative to the adatom cell, i.e., with three adatoms per cell. This structure lies 0.2~eV per adatom below the high-symmetry structure with the same slab thickness.
For this total-energy comparison, we used a $6\times6\times1$ $\mathbf{k}$-point mesh for the $\sqrt{3}\times\sqrt{3}$ supercell, which gives a sampling density comparable to that of the $12\times12\times1$ mesh for the high-symmetry cell ($12/\sqrt{3}\approx 6.9$). Apart from the slab thickness and the $\mathbf{k}$-point mesh, all parameters were kept unchanged.
Finally, we computed the phonon frequencies of the $\sqrt{3}\times\sqrt{3}$ reconstruction at $\Gamma$, onto which the $K$ point of the high-symmetry cell is folded. All frequencies are real, showing that the reconstruction removes the instability found at $K$ for the high-symmetry structure and that the reconstruction is stable against all distortions compatible with its $\sqrt{3}\times\sqrt{3}$ periodicity.
The reconstruction is shown in the B@B panel of Fig.~\ref{phonons}, where black lines mark the $\sqrt{3}\times\sqrt{3}$ unit cell containing three adatoms: two adatoms (red) remain at their T$_4$ sites, while the third (blue) relaxes from its T$_4$ site toward one of the neighboring H$_3$ sites.

These results are based on $T=0$ total energies and harmonic phonons at semi-local DFT level, which do not fully capture electronic correlation and anharmonic effects.
A useful comparison is the closely related Sn/Ge(111) system, whose reconstruction has the same periodicity and has been attributed to a surface soft phonon~\cite{perez2001softphonon}.
This phase is stable only at low temperatures: above $T_c\approx\SI{220}{\kelvin}$, the adatoms form an undistorted lattice on average~\cite{carpinelli1997surface,floreano2001orderdisorder}.
Whether, and up to which temperature, the B@B lattice reconstructs must therefore ultimately be settled by experiment.
\clearpage
\section{Surface and band symmetry}
\begin{figure}[h]
\includegraphics[width=\columnwidth]{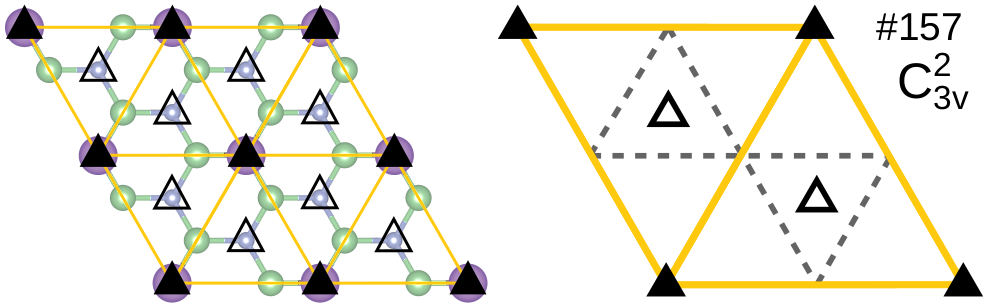}
\caption{C$_{3v}^2$ symmetry of the adatom-covered B-terminated surface. Left: Top view of $2\times 2$ units of the surface unit cell with the symmetry elements indicated. For clarity, only the uppermost BN bilayer (B: green, N: lavender) and the adatoms (purple) are shown. The T$_4$ adatom position coincides with a threefold rotation axis (black filled triangles). Neighboring adatoms are connected by mirror planes perpendicular to the surface (solid gold lines). Additional threefold rotation axes (black empty triangles) coincide with the topmost N sites, which lie at the center of three surrounding adatoms. Right: Schematic of space group no.~157 (C$_{3v}^2$) with the same symbols as on the left, plus additional glide planes (dashed grey lines). Adapted from the Supplemental Material of Ref.~\cite{menke2025}.}
\label{fig:space-group}
\end{figure}

\begin{figure}[h]
\includegraphics[width=0.98\columnwidth]{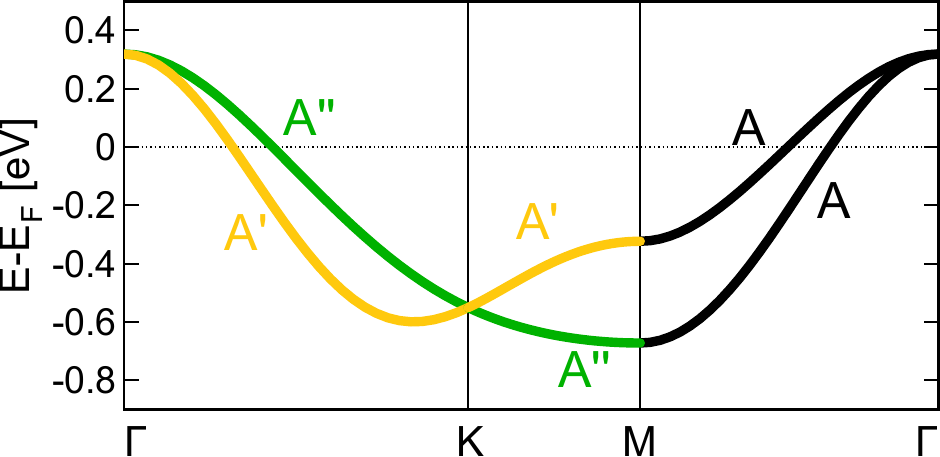}
\caption{\label{fig:band_sym} Symmetry of the two correlated bands, with Ga@B as a representative example. Along $\Gamma$--$K$--$M$, the little group $C_s$ has two irreducible representations, $A'$ (gold) and $A''$ (green), which are even and odd, respectively, under the vertical mirror plane $\sigma_v$ that protects the Dirac points in the absence of SOC. Along $M$--$\Gamma$, the little group reduces to $C_1$ with the single irreducible representation $A$ (black).}
\end{figure}

Figure~\ref{fig:space-group} shows the symmetry elements of the B-termination of cBN(111) with adatoms at the T$_4$ adsorption position (the N-termination is analogous). Highlighted are the vertical mirror planes $\sigma_v$ containing the nearest-neighbor adatom bonds, which protect the Dirac points in the absence of SOC.
Figure~\ref{fig:band_sym} shows the corresponding classification of the two correlated bands by the irreducible representations of their little group along the $\mathbf{k}$-path: along $\Gamma$--$K$--$M$, the bands are either even ($A'$) or odd ($A''$) under $\sigma_v$, so that the bands crossing at $K$ cannot hybridize.
\section{Wannier models}
The Wannier functions defining the correlated subspace (see main text) were obtained by projecting the isolated adatom bands onto atomic orbitals centered on the adatom: $p_z$ for the one-band systems ($z$ along the surface normal), $p_x$ and $p_y$ for the planar $p$-type doublet of group-III@B, and $d_{x^2-y^2}$ and $d_{xy}$ for the $d$-type doublet of TM@B, with $x$ along a nearest-neighbor bond.
Since the projected Wannier functions are already well localized, no spread minimization was performed.
The resulting basis thus retains the transformation properties of the projection orbitals and is adapted to the $C_{3v}$ site symmetry of the adatom. For the doublets, the two real Wannier functions form the partners of the irreducible representation $e$, one even and one odd under a vertical mirror plane $\sigma_v$.
Eigenstates of the threefold rotation would require the complex combinations $(|1\rangle \mp i|2\rangle)/\sqrt{2}$, so that the individual real Wannier functions break the $C_3$ symmetry, whereas all quantities computed in this basis preserve it.
\\

\begin{figure}[t]
\includegraphics[width=0.98\columnwidth]{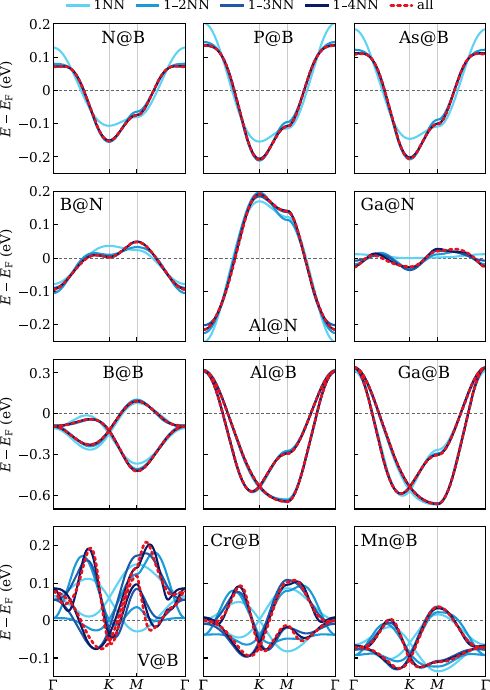}
\caption{\label{fig:TB_range} Wannier bands with hoppings truncated after the first (1NN) to fourth (1--4NN) neighbor shell (light to dark blue), compared with the full Wannier Hamiltonian (all, red dotted). Rows as in Fig.~\ref{fig:dft_all}.}
\end{figure}

In this basis, the Wannier Hamiltonian reads $H(\mathbf{k})=\sum_{\mathbf{R}}T(\mathbf{R})\,e^{i\mathbf{k}\cdot\mathbf{R}}$ with $[T(\mathbf{R})]_{mm'}=\langle
m\mathbf{0}|H|m'\mathbf{R}\rangle$ and $T(\mathbf{0})=(\varepsilon-E_\text{F})\openone$, where
$\mathbf{R}=n_1\mathbf{a}_1+n_2\mathbf{a}_2$ runs over the triangular adatom lattice
($|\mathbf{a}_{1,2}|=a=\SI{4.44}{\angstrom}$, $\angle(\mathbf{a}_1,\mathbf{a}_2)=120^\circ$, $\mathbf{a}_1\parallel x$).
Tables~\ref{tab:hopping_1band} and \ref{tab:hopping_2band} list the hoppings of the first four neighbor shells ($|\mathbf{R}|=a$, $\sqrt{3}a$, $2a$, $\sqrt{7}a$).
  \begin{table}[b]
  \caption{\label{tab:hopping_1band}Onsite energy and hoppings $t_n$ of shell $n$ (in meV) of the one-band Wannier models.}
  \begin{ruledtabular}
  \begin{tabular}{lrrrrr}
   & $\varepsilon-E_\text{F}$ & $t_1$ & $t_2$ & $t_3$ & $t_4$ \\ \hline
  N@B  & $-28$ & $26.104$ & $-8.078$  & $-1.576$ & $0.143$ \\
  P@B  & $-35$ & $39.601$ & $-9.488$  & $-1.941$ & $0.262$ \\
  As@B & $-36$ & $36.613$ & $-10.271$ & $-1.956$ & $0.113$ \\
  B@N  & $-1$  & $-12.766$ & $-4.510$ & $2.568$  & $-0.320$ \\
  Al@N & $30$  & $-46.932$ & $4.428$  & $3.919$  & $-1.108$ \\
  Ga@N & $4$   & $1.330$  & $-5.070$  & $1.787$  & $-1.409$ \\
  \end{tabular}
  \end{ruledtabular}
  \end{table}
For the one-band systems, the $a_1$ orbital is invariant under $C_{3v}$, so all vectors of shell $n$ carry the same hopping $t_n$.
For the two-band systems, the hopping matrices of symmetry-equivalent vectors are related by the transformation of the $e$ doublet,
\begin{equation}
\begin{split}
  T(C_3\mathbf{R})      &= D\,T(\mathbf{R})\,D^{T},\\
  T(\sigma_v\mathbf{R}) &= \tau_z\,T(\mathbf{R})\,\tau_z,\\
  T(-\mathbf{R})        &= T(\mathbf{R})^{T},
\end{split}
\label{eq:TB_sym}
\end{equation}
where $C_3$ maps $\mathbf{a}_1\to\mathbf{a}_2\to-\mathbf{a}_1-\mathbf{a}_2$, $\sigma_v$ is the mirror containing
$\mathbf{a}_1$, $\tau_z=\mathrm{diag}(1,-1)$ reflects the even ($p_x$, $d_{x^2-y^2}$) and odd ($p_y$, $d_{xy}$)
Wannier functions, and $D_{\pm}=\tfrac12\bigl(\begin{smallmatrix}-1&\mp\sqrt3\\\pm\sqrt3&-1\end{smallmatrix}\bigr)$ is
  a rotation by $\pm120^\circ$, with $D=D_{+}$ for $(p_x,p_y)$ and $D=D_{-}$ for $(d_{x^2-y^2},d_{xy})$, which rotate
  with twice the angle.
  Consequently, $T(\mathbf{a}_1)$ and $T(2\mathbf{a}_1)$ are diagonal, and one representative matrix per shell
  determines the model.
  With four shells, the models reproduce the bandwidths of the full Wannier bands to within 2$\%$ (see Fig.~\ref{fig:TB_range}).

\section{Nearest-neighbor interaction}
To estimate the nearest-neighbor interaction $V_{nn}$, we performed additional cRPA calculations for V/Cr/As@B as representative systems, using rectangular supercells containing two adatoms.
To keep these calculations feasible, the numerical parameters were reduced relative to those for the one-adatom cell given in the main text: we used a \SI{500}{\eV} energy cutoff, a vacuum region of \SI{16}{\angstrom}, a $7\times4\times1$ $\mathbf{k}$-point mesh, and 800 empty bands (states up to $\approx\SI{39}{\eV}$ above
$E_\text{F}$).
For the two-band systems, $V_{nn}$ is the average nearest-neighbor Coulomb interaction. The individual orbital- and bond-resolved values differ slightly because the Wannier functions individually break the threefold rotational symmetry (see prior section).
To benefit from the cancellation of systematic convergence errors, we normalize $V_{nn}$ by the on-site $U$ obtained in the same supercell calculation, which gives $V_{nn}/U=0.42$, $0.38$, and $0.37$ for V@B, Cr@B, and As@B, respectively.

\begin{table*}
  \caption{\label{tab:hopping_2band}Onsite energy $\varepsilon-E_\text{F}$ and hopping matrices $T(\mathbf{R}_n)$ (both in meV) of the two-band Wannier
  models for one representative vector $\mathbf{R}_n$ per shell; the header gives $|\mathbf{R}_n|$ and the number of vectors in the shell. All other
  matrices follow from Eq.~(\ref{eq:TB_sym}).}
  \begin{ruledtabular}
  \begin{tabular}{lccccc}
   & $\varepsilon-E_\text{F}$ & 1NN & 2NN & 3NN & 4NN \\
   & & $T(\mathbf{a}_1)$ & $T(\mathbf{a}_1+2\mathbf{a}_2)$ & $T(2\mathbf{a}_1)$ & $T(3\mathbf{a}_1+\mathbf{a}_2)$ \\
   & & $a$ (6) & $\sqrt{3}\,a$ (6) & $2a$ (6) & $\sqrt{7}\,a$ (12) \\ \hline
  B@B  & $-135$ & $\left(\begin{array}{@{}rr@{}}-50.843 & 0 \\ 0 & 60.436\end{array}\right)$ & $\left(\begin{array}{@{}rr@{}}-4.449 & 3.784 \\ -3.784 &
  11.612\end{array}\right)$ & $\left(\begin{array}{@{}rr@{}}-3.637 & 0 \\ 0 & -0.219\end{array}\right)$ & $\left(\begin{array}{@{}rr@{}}0.471 & -0.201 \\
  -0.180 & 0.292\end{array}\right)$ \\[8pt]
  Al@B & $-261$ & $\left(\begin{array}{@{}rr@{}}139.544 & 0 \\ 0 & 50.956\end{array}\right)$ & $\left(\begin{array}{@{}rr@{}}2.103 & 1.615 \\ -1.615 &
  1.042\end{array}\right)$ & $\left(\begin{array}{@{}rr@{}}-3.279 & 0 \\ 0 & -0.852\end{array}\right)$ & $\left(\begin{array}{@{}rr@{}}0.931 & -0.186 \\
  -0.258 & 0.285\end{array}\right)$ \\[8pt]
  Ga@B & $-272$ & $\left(\begin{array}{@{}rr@{}}147.779 & 0 \\ 0 & 47.421\end{array}\right)$ & $\left(\begin{array}{@{}rr@{}}-0.171 & 3.408 \\ -3.408 &
  8.929\end{array}\right)$ & $\left(\begin{array}{@{}rr@{}}-1.212 & 0 \\ 0 & -0.617\end{array}\right)$ & $\left(\begin{array}{@{}rr@{}}0.624 & -0.403 \\
  -0.062 & 0.285\end{array}\right)$ \\[8pt]
  V@B  & $62$   & $\left(\begin{array}{@{}rr@{}}-21.884 & 0 \\ 0 & 22.895\end{array}\right)$ & $\left(\begin{array}{@{}rr@{}}1.679 & 11.696 \\ -11.696 &
  -21.011\end{array}\right)$ & $\left(\begin{array}{@{}rr@{}}16.281 & 0 \\ 0 & 0.101\end{array}\right)$ & $\left(\begin{array}{@{}rr@{}}6.662 & -4.119 \\
  -7.100 & -1.605\end{array}\right)$ \\[8pt]
  Cr@B & $2$    & $\left(\begin{array}{@{}rr@{}}-20.975 & 0 \\ 0 & 21.199\end{array}\right)$ & $\left(\begin{array}{@{}rr@{}}-1.651 & 6.112 \\ -6.112 &
  -11.967\end{array}\right)$ & $\left(\begin{array}{@{}rr@{}}10.541 & 0 \\ 0 & -0.793\end{array}\right)$ & $\left(\begin{array}{@{}rr@{}}2.924 & -1.476 \\
  -2.399 & -0.103\end{array}\right)$ \\[8pt]
  Mn@B & $-61$  & $\left(\begin{array}{@{}rr@{}}-19.650 & 0 \\ 0 & 18.369\end{array}\right)$ & $\left(\begin{array}{@{}rr@{}}-3.763 & 2.933 \\ -2.933 &
  -5.493\end{array}\right)$ & $\left(\begin{array}{@{}rr@{}}6.691 & 0 \\ 0 & -1.684\end{array}\right)$ & $\left(\begin{array}{@{}rr@{}}1.395 & -0.507 \\
  -0.351 & 0.378\end{array}\right)$ \\
  \end{tabular}
  \end{ruledtabular}
  \end{table*}
\section{Analytic continuation}
\label{sec:app-ac}
The DMFT spectra require the self-energy on the real-frequency axis [Eqs.~(\ref{eq:spectral}) and (\ref{eq:green}) of the main text], whereas TRIQS/\textsc{cthyb} provides $\Sigma(i\omega_n)$ on the Matsubara axis. Since $\Sigma$ itself is not a normalized spectral function, we continue the auxiliary Green's function~\cite{Kraberger2017}
\begin{equation}
\label{eq:GAUX}
G_{\mathrm{aux}}(i\omega_n) = \left[i\omega_n + \mu_{\mathrm{aux}} - \Sigma(i\omega_n)\right]^{-1}.
\end{equation}
For any real $\mu_{\mathrm{aux}}$, $G_{\mathrm{aux}}$ has the analytic structure of a single-level Green's function with a non-negative spectral function of unit weight. It can therefore be continued with the well-established Maximum Entropy (MaxEnt) method, which is based on Bayesian statistics and reconstructs the real-frequency spectrum from a default model by maximizing its posterior probability~\cite{Skilling1984,Bryan1990}. We use the OmegaMaxEnt~\cite{bergeron2016algorithms} program.
The input self-energies were obtained in separate measurement runs after DMFT convergence, with the chemical potential and the Weiss field kept fixed. Compared to the self-consistency runs, these use fewer Monte Carlo cycles per measurement (\num{2e6} warm-up and \num{5e6} measurement cycles) but more than 1000 measurements per system, which provides reliable error estimates for MaxEnt.
The real-axis self-energy follows by inverting Eq.~\eqref{eq:GAUX},
\begin{equation}
\label{eq:GAUX2}
\Sigma(\omega) = \omega + \mu_{\mathrm{aux}} - G^{-1}_{\mathrm{aux}}(\omega).
\end{equation}
For an exact continuation, $\Sigma(\omega)$ would not depend on $\mu_{\mathrm{aux}}$. In practice it does, owing to the limited resolution of MaxEnt. We therefore continue $G_{\mathrm{aux}}$ for several $\mu_{\mathrm{aux}}$ and average the resulting real-frequency quantities. The shaded envelopes in the main-text figures and the error bars in Table~\ref{tab:overview} denote the standard deviation over $\mu_{\mathrm{aux}}$, i.e., the systematic uncertainty of the continuation.

\section{Atomic limit of the local spin susceptibility}
\label{sec:app-atomic}
Here we derive the atomic-limit values of $\chi_\text{loc}(\tau=0)$ quoted in the main text.

\paragraph{Instantaneous response.}
At $\tau=0$, Eq.~(\ref{eq:chi}) of the main text reduces to an equal-time expectation value,
\begin{equation}
  \chi_\text{loc}(\tau=0)=g^2\mu_B^2\,\big\langle (S_z^\text{tot})^2\big\rangle,
  \qquad S_z^\text{tot}=\sum_i S_z^i .
  \label{eq:supp-chi0}
\end{equation}
It measures the local moment present at a given instant, before it is screened by the itinerant electrons at larger $\tau$, and is therefore determined by the occupation probabilities of the local many-body states alone.

\paragraph{Atomic configurations.}
For $U\gg W$, these local states are the eigenstates of the isolated adatom with $N$ electrons in the correlated orbitals and the Hubbard-Kanamori interaction of the main text. Each of them has a definite total spin $S$, i.e., it is an eigenstate of $(\mathbf{S}^\text{tot})^2$ with eigenvalue $S(S+1)$ and belongs to a $(2S+1)$-fold degenerate multiplet ($S_z^\text{tot}=-S,\dots,S$).
In the one-band systems, the empty ($N=0$) and doubly occupied ($N=2$, $\uparrow\downarrow$) site have $S=0$, and the singly occupied site has $S=\tfrac12$.
In the two-band systems, the closed shells $N=0$ and $N=4$ have $S=0$, and one electron ($N=1$) or one hole ($N=3$) in the doublet gives $S=\tfrac12$. The six $N=2$ states are split by Hund's coupling ($U'=U-2J_\mathrm{H}$) into
\begin{equation}
\begin{split}
  E_{2_\mathrm{T}}&=U-3J_\mathrm{H}\quad(S=1,\ \times3),\\
  E_{2_\mathrm{S}}&=U-J_\mathrm{H}\ (\times2),\ \ U+J_\mathrm{H}\ (\times1),
\end{split}
  \label{eq:N2}
\end{equation}
where the high-spin triplet $2_\mathrm{T}$ has one electron in each orbital, $|{\uparrow},{\uparrow}\rangle$, $|{\downarrow},{\downarrow}\rangle$ and $(|{\uparrow},{\downarrow}\rangle+|{\downarrow},{\uparrow}\rangle)/\sqrt2$, and the singlets $2_\mathrm{S}$ ($S=0$) are the inter-orbital singlet and two combinations of doubly occupied orbitals. Since $2J_\mathrm{H}\geq\SI{0.48}{\eV}\gg k_\mathrm{B}T\approx\SI{0.01}{\eV}$, only the triplet is populated.

\paragraph{Single site.}
In the paramagnetic state, all ${2S+1}$ states of a multiplet are equally populated, so that $\langle(S_z^\text{tot})^2\rangle=\langle(\mathbf{S}^\text{tot})^2\rangle/3=S(S+1)/3$ and, with $g=2$,
\begin{equation}
  \chi_S=\tfrac43\,S(S+1)\,\mu_B^2
  =0,\ 1,\ \tfrac83~\mu_B^2
  \quad\text{for}\quad S=0,\ \tfrac12,\ 1.
  \label{eq:supp-chiS}
\end{equation}
For the triplet, each orbital contributes $\langle(S_z^a)^2\rangle=\langle(S_z^b)^2\rangle=\tfrac14$, and the remaining $\tfrac23-\tfrac12=\tfrac16$ is the inter-orbital correlation $2\langle S_z^aS_z^b\rangle$, which vanishes for two uncorrelated $S=\tfrac12$ spins (main text).

\paragraph{Non-integer filling.}
At $n=n_\mathrm{ref}\pm\delta$, the lowest atomic energy is reached if only the ground states of the two adjacent integer fillings are occupied, since any other charge state costs at least the charge gap ($U$ in the one-band systems, $U-3J_\mathrm{H}$ or $U+J_\mathrm{H}$ in the two-band systems). A fraction $1-\delta$ of the sites is then in the $n_\mathrm{ref}$ ground state and a fraction $\delta$ in the $n_\mathrm{ref}\pm1$ ground state, which gives
\begin{equation}
  \chi_\text{at}(\tau=0)=(1-\delta)\,\chi_{S(n_\mathrm{ref})}+\delta\,\chi_{S(n_\mathrm{ref}\pm1)} .
  \label{eq:supp-chimix}
\end{equation}
Table~\ref{tab:chi_atomic} lists the resulting values for ${\delta=0.25}$. For the doped Cr@B series, Eq.~\eqref{eq:supp-chimix} gives ${\chi_\text{at}(\tau=0)=\big(\tfrac83-\tfrac53\delta\big)~\mu_B^2}$, from $\tfrac83~\mu_B^2$ at $n=2$ to $2.25~\mu_B^2$ at $n=2.25$, which the DMFT values match within $0.5\%$ at every filling (Table~\ref{tab:chi_cr_doped}).
The DMFT values fall below Eq.~\eqref{eq:supp-chimix} by at most $0.04~\mu_B^2$ (Table~\ref{tab:chi_atomic}), as the finite bandwidth admixes further charge states.

\begin{table}[t]
\caption{Atomic ground states $(N,S)$ (electron number, spin) of the two charge states mixed at filling $n$, with their fractions, the resulting atomic value $\chi_\text{at}(\tau=0)$ of Eq.~\eqref{eq:supp-chimix} for $\delta=0.25$, and the DMFT values $\chi_\text{loc}(\tau=0)$ (both in $\mu_B^2$).}
\label{tab:chi_atomic}
\begin{tabular*}{\linewidth}{@{\extracolsep{\fill}}l c c c c c@{}}
\toprule
systems & $n$ & $(N,S)_{1-\delta}$ & $(N,S)_{\delta}$ & $\chi_\text{at}$ & $\chi_\text{loc}$ \\
\midrule
one-band @B     & $1+\delta$ & $(1,\tfrac12)$ & $(2,0)$ & $0.75$ & \multirow{2}{*}{$0.71$--$0.75$} \\
one-band @N     & $1-\delta$ & $(1,\tfrac12)$ & $(0,0)$ & $0.75$ & \\
B/Al/Ga/Mn@B    & $3+\delta$ & $(3,\tfrac12)$ & $(4,0)$ & $0.75$ & $0.71$--$0.75$ \\
V@B             & $1+\delta$ & $(1,\tfrac12)$ & $(2,1)$ & $1.42$ & $1.41$ \\
Cr@B            & $2+\delta$ & $(2,1)$ & $(3,\tfrac12)$ & $2.25$ & $2.24$ \\
\bottomrule
\end{tabular*}
\end{table}

\begin{table}[t]
  \caption{Instantaneous response along the doped Cr@B series of Sec.~\ref{sec:cr_doped}. Atomic value
  $\chi_\text{at}(\tau=0)=(1-\delta)\tfrac83+\delta$ of Eq.~\eqref{eq:supp-chimix} for a mixture of $S=1$ ($N=2$) and
  $S=\tfrac12$ ($N=3$) sites and DMFT value $\chi_\text{loc}(\tau=0)$ (in $\mu_B^2$).}
  \label{tab:chi_cr_doped}
  \begin{tabular*}{\linewidth}{@{\extracolsep{\fill}}c c S[table-format=1.2] S[table-format=1.2]@{}}
  \toprule
  $n$ & $\delta$ & {$\chi_\text{at}(\tau=0)$} & {$\chi_\text{loc}(\tau=0)$} \\
  \midrule
  $2.00$ & $0.00$ & 2.67 & 2.66 \\
  $2.05$ & $0.05$ & 2.58 & 2.57 \\
  $2.10$ & $0.10$ & 2.50 & 2.49 \\
  $2.15$ & $0.15$ & 2.42 & 2.41 \\
  $2.20$ & $0.20$ & 2.33 & 2.32 \\
  $2.25$ & $0.25$ & 2.25 & 2.24 \\
  \bottomrule
\end{tabular*}
\end{table}
\section{Hubbard-I atomic limit for one-band systems}
\label{sec:supp-hubbardI}
Here we derive the limiting value $Z^{-1}\to2/(1+\delta)$ and the position of the self-energy pole quoted in the main text for the one-band systems at fillings $n=1\pm\delta$ ($\delta=0.25$).

\paragraph{Spectral weight of the band at $E_\mathrm{F}$.}
In the atomic limit ($U\gg W$) and for $T\ll U$, each site is either singly occupied or, depending on the doping direction, doubly occupied or empty. For electron doping ($n=1+\delta$), a fraction $1-\delta$ of the sites is singly and a fraction $\delta$ doubly occupied. In the paramagnetic state, the spin of a singly occupied site is $\uparrow$ or $\downarrow$ with equal probability.
The band to which $E_\mathrm{F}$ is pinned is formed by the transitions between these two configurations. For a single orbital, each transition leads to exactly one final state, so its spectral weight per spin $\sigma$ is the probability that the transition is possible: adding a $\sigma$ electron requires a singly occupied site with spin $\bar\sigma$ [probability $(1-\delta)/2$], removing one requires a doubly occupied site (probability $\delta$). Hence
\begin{equation}
  w=\frac{1-\delta}{2}+\delta=\frac{1+\delta}{2}=n_{\bar\sigma}.
  \label{eq:supp-w1b}
\end{equation}
The remaining weight $1-w=(1-\delta)/2$ belongs to the transitions between empty and singly occupied sites, which form the satellite at $\omega\approx-U$.
For hole doping ($n=1-\delta$), the same counting with empty instead of doubly occupied sites yields again $w=(1+\delta)/2=1-n_{\bar\sigma}$, now with the satellite at $\omega\approx+U$.

\paragraph{Hubbard-I approximation.}
The Hubbard-I approximation~\cite{Hubbard1963} treats the hopping as a perturbation of the atom and is exact in the atomic limit. The self-energy is taken to be that of the isolated atom,
\begin{equation}
  \Sigma(\omega)=U n_{\bar\sigma}
  +\frac{n_{\bar\sigma}(1-n_{\bar\sigma})\,U^2}{\omega+\mu-(1-n_{\bar\sigma})\,U},
  \label{eq:supp-sigmaHI}
\end{equation}
and inserted into Eq.~(\ref{eq:green}) of the main text. Equivalently, $G(\mathbf{k},\omega)=[G_\mathrm{at}^{-1}(\omega)-\varepsilon(\mathbf{k})]^{-1}$ with the atomic Green's function
\begin{equation}
  G_\mathrm{at}(\omega)=\frac{1-n_{\bar\sigma}}{\omega+\mu}+\frac{n_{\bar\sigma}}{\omega+\mu-U}.
\end{equation}
For electron doping, $E_\mathrm{F}$ lies in the upper Hubbard band ($\mu\simeq U$). Near $E_\mathrm{F}$, only the pole carrying the weight $w=n_{\bar\sigma}$ contributes, $G_\mathrm{at}(\omega)\simeq w/(\omega-E_+)$ with $E_+=U-\mu$ its position relative to $E_\mathrm{F}$, so that
\begin{equation}
  G(\mathbf{k},\omega)\simeq\frac{w}{\omega-E_+-w\,\varepsilon(\mathbf{k})}.
  \label{eq:supp-GHI}
\end{equation}
The band thus carries the weight $w$, and its dispersion is the noninteracting one scaled by the same factor, ${\varepsilon(\mathbf{k})\to w\,\varepsilon(\mathbf{k})}$, while $E_+$ only shifts it rigidly. The same result follows directly from the slope of Eq.~\eqref{eq:supp-sigmaHI} at $\omega=0$:
\begin{equation}
  Z^{-1}=1-\left.\frac{\partial\,\real\Sigma}{\partial\omega}\right|_{\omega=0}
  =1+\frac{1-n_{\bar\sigma}}{n_{\bar\sigma}}=\frac{1}{w}=\frac{2}{1+\delta},
  \label{eq:supp-ZHI}
\end{equation}
i.e., $Z^{-1}=1.60$ at $\delta=0.25$. For hole doping ($\mu\simeq0$), the lower Hubbard band hosts $E_\mathrm{F}$ and one obtains $Z^{-1}=1/(1-n_{\bar\sigma})=1/w$, the same value.
In contrast to the Mott transition at half filling, where $Z^{-1}$ diverges, $Z^{-1}$ therefore stays finite for $U/W\to\infty$ at non-integer filling. The value depends only on $\delta$, not on $U$, $W$, or the shape of $\varepsilon(\mathbf{k})$.

\paragraph{Self-energy pole.}
Eq.~\eqref{eq:supp-sigmaHI} has a single pole at $\omega_\mathrm{res}=(1-n_{\bar\sigma})U-\mu$. With $\mu\simeq U$ (electron doping) or $\mu\simeq0$ (hole doping), this gives
\begin{equation}
  |\omega_\mathrm{res}|=w\,U=0.625\,U,
\end{equation}
on the satellite side, between the QP band at $E_\mathrm{F}$ and the satellite at $\pm U$. The hopping shifts $\mu$ within the band and hence $\omega_\mathrm{res}$ only by an amount of order $W$. The pole position is thus set by the atomic energy scale, in agreement with the observed $|\omega_\mathrm{res}|=(0.54$--$0.57)\,U$ for all one-band systems irrespective of $U/W$ (main text).

\paragraph{Validity.}
Hubbard-I neglects the coherence of the QPs, i.e., the dynamical screening of the local moments by the itinerant electrons. It is therefore exact only at temperatures well above the coherence scale, which is proportional to the bandwidth $W$. At fixed $T$, a large $U/W$ is thus necessary but not sufficient to reach the limit of Eq.~\eqref{eq:supp-ZHI}: the systems with the narrowest bands (Ga@N, B@N) come closest to it, whereas $Z^{-1}$ increases with $W$ for the others (main text).
\section{Satellites of the two-band systems}
\label{sec:supp-sat2b}

\paragraph{Atomic lines.}
In the electron-doped one-band systems, the only atomic transition besides the one forming the QP peak is the removal $1\rightarrow0$ (satellite at $-U$), since a doubly occupied site cannot accept a further electron. In the two-band systems, additional transitions appear for two reasons.
First, for $n_\mathrm{ref}<3$ the $n_\mathrm{ref}{+}1$ configuration is not a closed shell, so a further electron can be added ($2\rightarrow3$ for $n_\mathrm{ref}=1$, $3\rightarrow4$ for $n_\mathrm{ref}=2$).
Second, Hund's coupling splits the two-electron configuration into a high-spin triplet ground state ($2_\mathrm{T}$) and excited singlets ($2_\mathrm{S}$) lying $2J_\mathrm{H}$ and $4J_\mathrm{H}$ above it [Eq.~\eqref{eq:N2}].
Adding an electron to a singly occupied site or removing one from a triply occupied site can therefore also end in a singlet. These $1\rightarrow2_\mathrm{S}$ and $3\rightarrow2_\mathrm{S}$ processes give lines in addition to the transitions into $2_\mathrm{T}$.
Table~\ref{tab:atomic_lines} lists these lines with their weights, which are set by the fractions $1-\delta$ and $\delta$ of sites in the $n_\mathrm{ref}$ and $n_\mathrm{ref}{+}1$ ground states and, for doubly occupied final states, by their degeneracy (Sec.~\ref{sec:app-atomic}).
In the insets of Fig.~\ref{fig:dmft_spectral}, their positions are marked by green dashed lines.

\begin{table}[t]
\caption{Atomic excitation energies $\omega\neq0$ (measured from $E_\mathrm{F}$) and weights
per spin-orbital for $\delta=0.25$, with the cRPA parameters of
Table~\ref{tab:overview}. Transitions are labeled by the electron number of the initial
and final atomic configuration; $2_\mathrm{T}$ ($2_\mathrm{S}$) denotes the two-electron
triplet (singlets), and $J\equiv J_\mathrm{H}$.}
\label{tab:atomic_lines}
\begin{tabular*}{\linewidth}{@{\extracolsep{\fill}}l c l l
                S[table-format=+1.2, retain-explicit-plus]
                S[table-format=1.2]@{}}
\toprule
system & $n_\mathrm{ref}$ & transition & line & {$\omega$ [eV]} & {weight} \\
\midrule
V@B  & 1 & $1\rightarrow0$             & $-(U{-}3J)$ & -0.60 & 0.19 \\
     &   & $1\rightarrow2_\mathrm{S}$  & $+2J$       & +0.48 & 0.19 \\
     &   & $1\rightarrow2_\mathrm{S}$  & $+4J$       & +0.97 & 0.09 \\
     &   & $2_\mathrm{T}\rightarrow3$  & $+(U{+}J)$  & +1.57 & 0.13 \\
\addlinespace
Cr@B & 2 & $2_\mathrm{T}\rightarrow1$  & $-(U{+}J)$  & -1.87 & 0.38 \\
     &   & $3\rightarrow2_\mathrm{S}$  & $-4J$       & -1.28 & 0.03 \\
     &   & $3\rightarrow2_\mathrm{S}$  & $-2J$       & -0.64 & 0.06 \\
     &   & $3\rightarrow4$             & $+(U{-}3J)$ & +0.59 & 0.06 \\
\addlinespace
B@B  & 3 & $3\rightarrow2_\mathrm{S}$  & $-(U{+}J)$  & -3.47 & 0.09 \\
     &   & $3\rightarrow2_\mathrm{S}$  & $-(U{-}J)$  & -2.93 & 0.19 \\
     &   & $3\rightarrow2_\mathrm{T}$  & $-(U{-}3J)$ & -2.38 & 0.28 \\
\addlinespace
Al@B & 3 & $3\rightarrow2_\mathrm{S}$  & $-(U{+}J)$  & -2.80 & 0.09 \\
     &   & $3\rightarrow2_\mathrm{S}$  & $-(U{-}J)$  & -2.32 & 0.19 \\
     &   & $3\rightarrow2_\mathrm{T}$  & $-(U{-}3J)$ & -1.84 & 0.28 \\
\addlinespace
Ga@B & 3 & $3\rightarrow2_\mathrm{S}$  & $-(U{+}J)$  & -2.81 & 0.09 \\
     &   & $3\rightarrow2_\mathrm{S}$  & $-(U{-}J)$  & -2.35 & 0.19 \\
     &   & $3\rightarrow2_\mathrm{T}$  & $-(U{-}3J)$ & -1.88 & 0.28 \\
\addlinespace
Mn@B & 3 & $3\rightarrow2_\mathrm{S}$  & $-(U{+}J)$  & -2.18 & 0.09 \\
     &   & $3\rightarrow2_\mathrm{S}$  & $-(U{-}J)$  & -1.40 & 0.19 \\
     &   & $3\rightarrow2_\mathrm{T}$  & $-(U{-}3J)$ & -0.62 & 0.28 \\
\bottomrule
\end{tabular*}
\end{table}
\paragraph{Satellite weight and mean energy.}
Far from $E_\mathrm{F}$, the individual lines are not resolved, since the incoherent excitations acquire a finite lifetime in the metallic state and the resolution of the analytic continuation decreases with $|\omega|$. We therefore characterize each satellite by its spectral weight and mean energy,
\begin{align}
w_\mathrm{sat} &= \int_\mathrm{sat}\!A(\omega)\,d\omega, \label{eq:wsat}\\
\langle\omega\rangle_\mathrm{sat} &= \frac{1}{w_\mathrm{sat}}\int_\mathrm{sat}\!\omega\,A(\omega)\,d\omega, \label{eq:omsat}
\end{align}
where $A(\omega)=-\frac{1}{2\pi}\sum_m\imag G_{mm}(\omega)$ is the orbital-averaged local spectral function (total weight $1$ per spin-orbital), and the integration covers the entire satellite, starting from the spectral minimum between QP peak and satellite.
Unlike the line shape, both quantities hardly depend on the details of the analytic continuation.
Their atomic counterparts are the total weight $w_\mathrm{at}$ and the weighted mean energy $\langle\omega\rangle_\mathrm{at}$ of the atomic lines on the same side of $E_\mathrm{F}$.

\paragraph{Comparison.}
Table~\ref{tab:sat2b} shows that in all two-band systems the mean energies of the satellites below $E_\mathrm{F}$, and for V@B also of the one above $E_\mathrm{F}$, agree with the atomic values to within \SI{0.1}{\eV}.
For B/Al/Ga@B ($J_{\mathrm{H}}/U\approx0.09$), the lines span only about a third of $U$ and merge into a single peak, whereas for Mn@B ($J_{\mathrm{H}}/U=0.22$) they spread over $4J_{\mathrm{H}}=\SI{1.56}{\eV}$, comparable to $U$, and the satellite becomes broad and asymmetric.
Above $E_\mathrm{F}$, the weak $3\rightarrow4$ line of Cr@B is not resolved and merges into the high-energy tail of the QP peak.
The satellite maxima show the same spread as the mean energies ($|\omega_\mathrm{sat}|/U=0.39$--$0.94$, Table~\ref{tab:overview}).
The weights $w_\mathrm{sat}$ reach $\gtrsim90\%$ of $w_\mathrm{at}$, except for the Fermi liquids Al/Ga@B ($\approx80\%$), where a self-energy resonance splits off a third peak at $\approx\SI{0.5}{\eV}$ that takes weight from the QP peak.

\begin{table}
\caption{Satellites of the two-band systems: mean energy $\langle\omega\rangle_\mathrm{sat}$ and spectral weight $w_\mathrm{sat}$ (per spin-orbital), compared with the weighted mean energy $\langle\omega\rangle_\mathrm{at}$ and the total weight $w_\mathrm{at}$ of the corresponding atomic lines of Table~\ref{tab:atomic_lines}; for V@B, satellites occur at both
$\omega<0$ (V@B$^-$) and $\omega>0$ (V@B$^+$).}
\label{tab:sat2b}
\small
\setlength{\tabcolsep}{4pt}
\begin{tabular*}{\linewidth}{@{\extracolsep{\fill}}l c
                  S[table-format=-1.1] S[table-format=-1.1]
                  S[table-format=1.2]  S[table-format=1.2]@{}}
  \toprule
   & & \multicolumn{2}{c}{$\langle\omega\rangle$ [eV]} & \multicolumn{2}{c}{$w$} \\
  \cmidrule(lr){3-4}\cmidrule(l){5-6}
  system & $n_\mathrm{ref}$ & {sat} & {at} & {sat} & {at} \\
  \midrule
  V@B$^-$ & 1 & -0.6 & -0.6 & 0.17 & 0.19 \\
  V@B$^+$ & 1 & 1.0 & 0.9 & 0.37 & 0.41 \\
  Cr@B & 2 & -1.6 & -1.7 & 0.46 & 0.47 \\
  B@B  & 3 & -2.8 & -2.7 & 0.51 & 0.56 \\
  Al@B & 3 & -2.2 & -2.2 & 0.45 & 0.56 \\
  Ga@B & 3 & -2.3 & -2.2 & 0.44 & 0.56 \\
  Mn@B & 3 & -1.1 & -1.1 & 0.52 & 0.56 \\
  \bottomrule
  \end{tabular*}
\end{table}

\paragraph{Doped Cr@B series.}
For the Cr@B series of Sec.~\ref{sec:cr_doped} ($n=2+\delta$, $\delta=0.05,\dots,0.25$), the atomic lines below $E_\mathrm{F}$ are the removal $2_\mathrm{T}\rightarrow1$ at $-(U+J_\mathrm{H})$ with weight $(1-\delta)/2$ and the removals $3\rightarrow2_\mathrm{S}$ at $-2J_\mathrm{H}$ and $-4J_\mathrm{H}$ with weights $\delta/4$ and $\delta/8$ per spin-orbital (cf.\ Table~\ref{tab:atomic_lines} for $\delta=0.25$). Their weighted mean energy and total weight are
\begin{equation}
\langle\omega\rangle_\mathrm{at}
=-\frac{4(1-\delta)(U+J_\mathrm{H})+8\delta J_\mathrm{H}}{4-\delta},\qquad
w_\mathrm{at}=\frac{4-\delta}{8}.
\label{eq:supp-cr-at}
\end{equation}
From $\delta=0.05$ to $0.25$, this gives $\langle\omega\rangle_\mathrm{at}=-1.83$ to $\SI{-1.67}{\eV}$ and $w_\mathrm{at}=0.494$ to $0.469$, compared with $\langle\omega\rangle_\mathrm{sat}=-1.83$ to $\SI{-1.64}{\eV}$ and $w_\mathrm{sat}=0.492$ to $0.461$ from DMFT, i.e., agreement within $2\%$ at every filling.

\section{AI usage}
Claude Code (Anthropic, model Claude Opus 5.5) was used to draft and polish parts of the text and to double check the Wannier symmetry, atomic-limit and Hubbard~I results in this Supplemental Material. The authors reviewed and revised all text and derivations and take full responsibility for the content.

\end{document}